\documentclass[longauth]{aa}  
\usepackage{graphicx}
\usepackage{txfonts}
\usepackage[normalem]{ulem}

\usepackage{tabularx}
\usepackage{scrextend}
\usepackage{lscape} 
\usepackage{bigstrut}

\usepackage{lineno} 

\usepackage{soul}
\usepackage{color}

\newcommand{\pkg}[1]{\texttt{#1}}
\newcommand{\salt}{\pkg{SALT2.4}}
\newcommand{\saltto}{\pkg{SALT2.1}}
\newcommand{\sep}{\pkg{SEP}}

\newcommand{\SNNGC}{SN2013bs}  
\newcommand{\SNPGC}{SN2013bt} 

\newcommand{\nSNe}{141} 
\newcommand{\nRed}{five} 
\newcommand{\nOH}{65} 
\newcommand{\nSNeUntargeted}{114} 
\newcommand{\nSNeYoung}{70} 
\newcommand{\nSNeOld}{71} 

\newcommand{\DY}{\ensuremath{\Delta_{Y}}}
\newcommand{\DM}{\ensuremath{\Delta_{M}}}
\newcommand{\PY}{\ensuremath{{\mathcal P}_Y}}
\newcommand{\PM}{\ensuremath{{\mathcal P}_{HM}}}

\newcommand{\Instru}[1]{\textsc{#1}}

\usepackage{natbib,twoopt}
\usepackage[breaklinks=true, draft]{hyperref} 
\bibpunct{(}{)}{;}{a}{}{,}             
\makeatletter
  \newcommandtwoopt{\citeads}[3][][]{\href{http://adsabs.harvard.edu/abs/#3}%
    {\def\hyper@linkstart##1##2{}%
     \let\hyper@linkend\@empty\citealp[#1][#2]{#3}}}
  \newcommandtwoopt{\citepads}[3][][]{\href{http://adsabs.harvard.edu/abs/#3}%
    {\def\hyper@linkstart##1##2{}%
     \let\hyper@linkend\@empty\citep[#1][#2]{#3}}}
  \newcommandtwoopt{\citetads}[3][][]{\href{http://adsabs.harvard.edu/abs/#3}%
    {\def\hyper@linkstart##1##2{}%
     \let\hyper@linkend\@empty\citet[#1][#2]{#3}}}
  \newcommandtwoopt{\citeyearads}[3][][]%
    {\href{http://adsabs.harvard.edu/abs/#3}
    {\def\hyper@linkstart##1##2{}%
     \let\hyper@linkend\@empty\citeyear[#1][#2]{#3}}}
\makeatother

\begin{document}

\title{Strong dependence of type Ia supernova standardization on the local
  specific star formation rate}

\titlerunning{Local~sSFR bias in SNe~Ia}

\author{M.~Rigault \inst{\ref{ipnl},\ref{lpc}} 
\and   V.~Brinnel \inst{\ref{berlin}} 
\and   G.~Aldering \inst{\ref{lbnl}}
\and   P.~Antilogus \inst{\ref{lpnhe}}
\and   C.~Aragon \inst{\ref{lbnl}}
\and   S.~Bailey \inst{\ref{lbnl}}
\and   C.~Baltay \inst{\ref{yale}} 
\and   K.~Barbary \inst{\ref{lbnl}} 
\and   S.~Bongard \inst{\ref{lpnhe}}
\and   K.~Boone \inst{\ref{lbnl},\ref{ucb}}
\and   C.~Buton \inst{\ref{ipnl}}
\and   M.~Childress \inst{\ref{southampton}}
\and   N.~Chotard \inst{\ref{ipnl}}
\and   Y.~Copin \inst{\ref{ipnl}}
\and   S.~Dixon \inst{\ref{lbnl}}
\and   P.~Fagrelius \inst{\ref{lbnl},\ref{ucb}}
\and   U.~Feindt \inst{\ref{okc}}
\and   D.~Fouchez \inst{\ref{cppm}}
\and   E.~Gangler \inst{\ref{lpc}}  
\and   B.~Hayden \inst{\ref{lbnl}}
\and   W.~Hillebrandt \inst{\ref{garching}}
\and   D.~A. Howell \inst{\ref{lcogt},\ref{ucsb}} 
\and   A.~Kim \inst{\ref{lbnl}}
\and   M.~Kowalski \inst{\ref{berlin},\ref{desy}}
\and   D.~Kuesters \inst{\ref{berlin}}
\and   P.-F.~Leget, \inst{\ref{lpc}}
\and   S.~Lombardo \inst{\ref{berlin}}
\and   Q.~Lin \inst{\ref{china}}
\and   J.~Nordin \inst{\ref{berlin}}
\and   R.~Pain \inst{\ref{lpnhe}}
\and   E.~Pecontal \inst{\ref{cral}}
\and   R.~Pereira \inst{\ref{ipnl}}
\and   S.~Perlmutter \inst{\ref{lbnl},\ref{ucb}}
\and   D.~Rabinowitz \inst{\ref{yale}} 
\and   K.~Runge \inst{\ref{lbnl}} 
\and   D.~Rubin \inst{\ref{lbnl},\ref{stsci}}
\and   C.~Saunders \inst{\ref{lpnhe}}
\and   G.~Smadja \inst{\ref{ipnl}} 
\and   C.~Sofiatti \inst{\ref{lbnl},\ref{ucb}}
\and   N.~Suzuki \inst{\ref{lbnl},\ref{ipmu}}
\and   S.~Taubenberger \inst{\ref{garching},\ref{eso}}
\and   C.~Tao \inst{\ref{cppm},\ref{china}}
\and   R.~C.~Thomas \inst{\ref{nersc}}\\
\textsc{The Nearby Supernova Factory}
}


\institute{
      Universit\'e de Lyon, F-69622, Lyon, France; Universit\'e de Lyon 1, Villeurbanne; 
      CNRS/IN2P3, Institut de Physique des Deux Infinis, Lyon. \label{ipnl}
\and 
    Université Clermont Auvergne, CNRS/IN2P3, Laboratoire de Physique
    de Clermont, F-63000 Clermont-Ferrand, France. \label{lpc} 
\and
    Institut fur Physik, Humboldt-Universitat zu Berlin,
    Newtonstr. 15, 12489 Berlin \label{berlin} 
\and 
    Physics Division, Lawrence Berkeley National Laboratory, 
    1 Cyclotron Road, Berkeley, CA, 94720 \label{lbnl}
\and 
    Laboratoire de Physique Nucl\'eaire et des Hautes \'Energies,
    Universit\'e Pierre et Marie Curie Paris 6, Universit\'e Paris Diderot Paris 7, CNRS-IN2P3, 
    4 place Jussieu, 75252 Paris Cedex 05, France \label{lpnhe}
\and 
    Department of Physics, Yale University, 
    New Haven, CT, 06250-8121 \label{yale}
\and 
    Department of Physics, University of California Berkeley,
    366 LeConte Hall MC 7300, Berkeley, CA, 94720-7300 \label{ucb}
\and 
    Department of Physics and Astronomy, University of Southampton,
    Southampton, Hampshire, SO17 1BJ, UK \label{southampton} 
\and 
    The Oskar Klein Centre, Department of Physics, AlbaNova, Stockholm
    University, SE-106 91 Stockholm, Sweden \label{okc}
\and 
    Aix Marseille Universit\'e, CNRS/IN2P3, CPPM UMR 7346, 13288,
    Marseille, France \label{cppm}
\and
    Max-Planck Institut f\"ur Astrophysik, Karl-Schwarzschild-Str. 1,
    85748 Garching, Germany \label{garching}
\and 
    Las Cumbres Observatory Global Telescope Network, 6740 Cortona
    Dr., Suite 102 Goleta, Ca 93117 \label{lcogt} 
\and 
    Department of Physics, University of California, Santa Barbara, CA
    93106-9530, USA \label{ucsb} 
\and
    Deutsches Elektronen-Synchrotron, D-15735 Zeuthen, Germany \label{desy}
\and 
    Tsinghua Center for Astrophysics, Tsinghua University, Beijing
    100084, China \label{china} 
\and 
    Centre de Recherche Astronomique de Lyon, Universit\'e Lyon 1,
    9 Avenue Charles Andr\'e, 69561 Saint Genis Laval Cedex,
    France \label{cral}
\and
    Space Telescope Science Institute, 3700 San Martin Drive,
    Baltimore, MD 21218 \label{stsci}
\and
    European Southern Observatory, Karl-Schwarzschild-Str. 2, 85748
    Garching, Germany \label{eso}
\and
    Computational Cosmology Center, Computational Research Division, Lawrence Berkeley National Laboratory, 
    1 Cyclotron Road MS 50B-4206, Berkeley, CA, 94720 \label{nersc}
\and
    Kavli Institute for the Physics and Mathematics of the Universe,
    University of Tokyo, 5-1-5 Kashiwanoha, Kashiwa, Chiba, 277-8583, Japan \label{ipmu}
}

\date{}

\abstract{

As part of an on-going effort to identify, understand and correct for 
astrophysics biases in the standardization of type Ia supernovae (SN~Ia) for
cosmology, we have statistically classified a large sample of nearby SNe~Ia
into those that are located in predominantly younger or older
environments. 
This classification is based on the specific star formation rate 
measured within a projected distance of 1~kpc from each SN location (LsSFR).
This is an important refinement compared to using the local star
formation rate directly, as it
provides a normalization for relative numbers of available SN
progenitors and is more robust against extinction by dust. 
We find that the SNe~Ia in predominantly younger environments
are $\Delta_{Y} = 0.163 \pm 0.029\,\mathrm{mag}$ ($5.7\,\sigma$) fainter
than those in predominantly older environments after
conventional light-curve standardization.
This is the strongest
standardized SN Ia
brightness systematic connected to the host-galaxy environment
measured to date.
The well-established step in standardized brightnesses between
SNe~Ia in hosts with lower or higher total stellar masses is
smaller, at
$\DM= 0.119 \pm 0.032\,\mathrm{mag}$ ($4.5\,\sigma$), for the
same set of SNe~Ia.  When fit simultaneously, the environment-age
offset remains very significant, with $\Delta_{Y} = 0.129 \pm
0.032$~mag ($4.0\,\sigma$), while the global stellar mass step is reduced to
$\Delta_M = 0.064 \pm 0.029$~mag ($2.2\,\sigma$). Thus, approximately
70\% of the variance from the stellar mass step is due to an underlying
dependence on environment-based progenitor age.  
Also, we verify that using the local star formation rate alone
is not as powerful as LsSFR at sorting SNe~Ia into brighter and fainter
subsets.
Standardization that only uses the SNe~Ia in younger
environments reduces the total
dispersion from $0.142\pm0.008\,\mathrm{mag}$ to $0.120 \pm 0.010\,\mathrm{mag}$.
We show that as environment-ages evolve with
redshift, a strong bias, especially on the measurement of the derivative
of the dark energy equation of state, can develop. Fortunately,
data that measure and correct for this effect using our local specific
star formation rate indicator, are likely to be available for many
next-generation SN~Ia cosmology experiments.
}


\maketitle

\section{Introduction}
\label{sec:introduction}

Empirically standardized type Ia supernovae (SNe~Ia) are powerful
cosmological distance indicators that enable us to trace the
expansion history of the Universe.  The SN~Ia
redshift-magnitude relation led to the discovery of the accelerated
expansion of the Universe \citep{perlmutter_1999, riess_1998}, which is
attributed to an elusive dark energy. This acceleration
has been confirmed with high precision
\citep{betoule_2014,planck_cosmo_2015}.
Since SNe~Ia directly probe the period when the expansion of the Universe
is driven by dark energy, they remain a key probe for
cosmology \citep{kim_2015}.  They are particularly powerful at
measuring the dark energy equation of state parameter $w$ and its
potential redshift evolution $w_a$ \citep{weinberg_2013, betoule_2014},
as well as directly deriving the Hubble constant, $H_0$
\citep{riess_2009,riess_2016}.
At present, the direct measurement of $H_0$ using SNe~Ia disagrees
significantly with extrapolation to the current epoch of cosmic
microwave background constraints \citep{planck_cosmo_2015, riess_2016}.
This could signify other new physics, but there still may be
systematic effects that are unaccounted for \citep{rigault_2015}.

The SN~Ia distance measurement technique relies on the ability of
determining SN luminosities over a wide range of redshifts in a consistent way.
The observed luminosity dispersion for normal SNe~Ia is
$\sim40$\%. This can be significantly reduced to $\sim15$\% when using
empirical relations between the SN lightcurve peak luminosity and
the lightcurve shape and color \citep{phillips_1993,
riess_1996,tripp_1998}.  The basic behavior underlying this
standardization is that fainter supernovae are redder, and brighten
and fade more quickly.  

\cite{fakhouri_2015} have subsequently shown that twin SNe~Ia,
which are pairs with very similar spectra at peak-luminosity, exhibit a
dispersion in luminosity below 8\%.  
This result was obtained using the
same data as lightcurve-based standardization, implying that the
15\% dispersion based on optical light curves is not random, but
rather correlated in some unknown manner so that the use of twin SNe~Ia
is able to cancel, at least at the level of measurement
  uncertainties.
Lower dispersion is also found at
near-infrared wavelengths; \cite{barone-nugent_2012} find a dispersion
of 9\%.  Here models suggest that astrophysical differences in SNe
should be lower than in optical bands \citep{kasen_2006}.
Taken together, these results motivate the search for clues concerning
the nature of the astrophysical differences that existing
standardization does not yet fully remove.

Constraints
on the true nature of SN~Ia progenitors remains limited.
\citealt{branch_1995},
\citealt{hillebrandt_niemeyer_2000}, \citealt{maoz_mannucci_2014},
and \citealt{maeda_terada_2016} provide comprehensive reviews of
potential explosion 
scenarios and some of their expected observational signatures.
Since the impact of progenitor properties, such as mass, age, or metallicity,
on the resulting standardized peak luminosity is not constrained
well enough by models so as to be applied anywhere near the precision required
for cosmological measurements,
effort has focused
on empirical studies beyond the direct measurement of the SNe~Ia light curves.

One such avenue that has proven productive has been the study of
correlations between SNe~Ia and their host-galaxy properties.  For
instance, a strong quantitative connection between the total stellar
mass and standardized SN brightnesses is now well established
\citep{kelly_2010, sullivan_2010, gupta_2011,childress_2013}.
Correlations of standardized brightnesses with host-galaxy stellar
age \citep{sullivan_2010,gupta_2011} and gas-phase metallicity
\citep{dandrea_2011,childress_2013} have been identified. In
addition, the distributions of light-curve widths used to
standardize SNe~Ia have been shown to change depending on the host
galaxy morphology \citep{hamuy_1996, hamuy_2000}, total stellar
mass \citep{neill_2009,sullivan_2010, lampeitl_2010,kelly_2010,gupta_2011},
stellar metallicity \citep{dandrea_2011, pan_2014}, global specific
star formation rate \citep[sSFR;][]{sullivan_2006}, stellar age
\citep{neill_2009,lampeitl_2010,childress_2013} and local star formation rate
\citep{rigault_2013}.

These results are clear evidence that host-galaxy properties and
variations in the progenitor populations are connected, and that
astrophysical biases remain after the usual lightcurve stretch and color
standardization. Determining the cause is complicated by the
fact that, for example, galaxy stellar mass simultaneously correlates with
stellar metallicity \citep[e.g.,][]{tremonti_2004} and stellar age
\citep[e.g.,][]{gallazzi_2005}, and morphology correlates with stellar
age as well.
Since relative host
galaxy stellar masses are straightforward to derive from deep
broadband imaging that accompanies modern SN search and follow-up,
a brightness step between
SNe~Ia with  host stellar masses on either side of
$\log(\mathrm{M_*/M_\odot})=10$ 
is now commonly included as a third standardization parameter
\citep{sullivan_2011,suzuki_2012,betoule_2014}.

However, galaxy stellar mass itself is unlikely to be
the root cause of the effect. Being stars, SN will have formed in
a group with other stars, having common ages and metallicities. As
discussed in \citet{rigault_2013} and \citet{rigault_2015}, such
groups initially have low velocity dispersions, which imply timescales
of $\sim300$~Myr to dispersion by a distance of 1~kpc. Even then,
most of the velocity is in the form of angular momentum, and so
those stars tend to oscillate around a mean galactocentric distance,
preserving their correlation with other nearby star over much longer
periods of time. By comparison, global properties for isolated
galaxies are primarily governed by the dark matter halo mass and
the amount of infalling gas. These factors correlate with some 
bulk properties of stars in a galaxy, but local correlations remain
the strongest. Such local coherence in stellar properties has
long been exploited in estimating relative ages of supernovae
\citep[e.g.,][]{moore_1973,vandyk_1992,bartunov_1994, aramyan_2016}.

An additional confounding factor is that SN~Ia observed brightnesses
are dimmer and their colors redder due to dust, and dust correlates
with many galaxy properties.  Measurement of galaxy properties,
such as light-weighted stellar mass and age, depend on modeling to
correct for dust, which is complicated by scattering.  Dust extinction
curves are found to vary between the Galaxy, the LMC and SMC,
suggesting the influence of metallicity or other differences in the
interstellar media of these galaxies. Metals are needed to build
dust grains and gas-phase metallicity correlates with total stellar
mass, inducing a correlation between the amount
\citep[e.g.,][]{brinchmann_2004,garn_2010,battisti_2016}, and
potentially, the properties of dust as a function of host-galaxy
stellar mass. Dust
should also correlate with age since shocks can destroy dust grains
and galactic winds can remove them.  Even the mean path length for
SN light to escape a galaxy depend on its size.
Care is therefore required in the measurement and interpretation
of SN host-galaxy environmental effects.

Even so, substantial progress has been made.
For instance, \cite{childress_2013} exploited the 
sharpness of the change in standardized SN  brightnesses on either
side of the transition at a total stellar mass of
$\log(M_*/M_\odot)\sim10$, finding that metallicity or dust
extinction change too smoothly with galaxy stellar mass to be the
primary driver. Only star-formation, which follows the ``main sequence''
of galaxy formation, shows a sufficiently sharp transition versus
galaxy stellar mass \citep{salim_2007,noeske_2007, elbaz_2007,daddi_2007},
and this transition occurs at the right global stellar mass to match
the SN data.

Another key constraint on progenitors has come from correlations
between SN~Ia rates and host-galaxy properties used to estimate
the delay time distribution (DTD), that is, the time from initial
formation of the progenitor system to explosion. Initial work
suggested a ``prompt'' subpopulation with a rate proportional to
the star-formation rate, plus a ``delayed'' subpopulation whose
rate is proportional to host-galaxy stellar mass \citep{mannucci_2005,
scannapieco_2006,sullivan_2006,aubourg_2008}. Studies using host
galaxy ages and evolution with redshift indicate a smoother DTD
falling roughly as $t^{-1}$ \citep[see][for a detailed review]{
maoz_mannucci_2014}.  
When this smooth distribution is convolved with the main sequence of
galaxy formation, a bimodal distribution 
of younger and older modes is expressed in the age distribution of
SNe~Ia \citep{childress_2014}. 
The young (aka. prompt) population is continuously renewed
by star formation and therefore its age distribution remains fixed
at young ages, whereas the mean age of the old/delayed distribution is
tied to the large numbers of stars formed in massive galaxies when
the universe was young.

In a further effort to quantify the connection between SN~Ia
progenitors, SN~Ia standardized brightnesses and star-formation/age,
several studies have pioneered the use of the host-galaxy region in the
immediate vicinity of SNe~Ia \citep{stanishev_2012, rigault_2013,
galbany_2014}.  Of these, the \citet{rigault_2013}
study was the first having a sample large enough for a
statistical analysis, with 82 Hubble-flow SNe~Ia, and 
used H$\alpha$ emission
within a projected radius of 1~kpc as a star-formation tracer. When
comparing the properties of the SNe~Ia from low- and high-star
forming regions it was found that SNe~Ia with high local star
formation are fainter after standardization, and significantly less
dispersed in brightness. \citet{rigault_2015}, who used a 2~kpc radius
aperture, and \cite{kelly_2015}, who used a 5~kpc aperture, confirmed
these results using GALEX UV imaging. 
\cite{jones_2015} replicated the study of \citet{rigault_2015},
though they then found a weaker effect when using a different light
curve fitter, adding some additional SNe~Ia, and applying additional
selection cuts to their sample.\footnote{After submission
of the current paper, both \cite{roman_2018} and \cite{kim_2018}
have confirmed the presence of a similar bias using alternative
local host-galaxy properties.}

Continuing the thread of our previous host analyses \citep{childress_2013,
rigault_2013, rigault_2015}, this paper aims at building an
astrophysically-motivated SN host analysis that would allow a more
direct interpretation of the sources of observed correlations.  Here
we focus on using the specific star formation rate indicator observed
at the SN location to trace the impact of the SN progenitor age on
the SN lightcurve parameters and standardized magnitudes. In
Section~\ref{sec:discussion_indicators} we further discuss the
importance of choosing an accurate galaxy indicator for SN host
analyses, showing that the local~sSFR (LsSFR)  is a natural parameter
to probe intrinsic SN variations.  The derivation of these parameters
is presented in Section~\ref{sec:data} and associated results are
given in Section~\ref{sec:results}. We provide comparisons with
previous results, and several cross-checks,
in Section~\ref{sec:discussion}. We then turn to the consequences
of our results for cosmology in Section~\ref{sec:cosmo_bias} and then
summarize and conclude in Section~\ref{sec:conclusion}.

\section{The LsSFR host-galaxy environment indicator}
\label{sec:discussion_indicators}

Studies of host-galaxy global properties often use the specific
star formation rate, sSFR, the star formation rate (SFR) normalized 
by the stellar mass, to rank galaxies by their relative star formation
activity. Here we extend this approach to the host-galaxy regions
projected on the sky in the vicinity of individual SNe~Ia. In the
context of SN~Ia progenitor models, the LsSFR will be
correlated with the relative numbers of young and old progenitor
systems. This is very appealing for the study of SN~Ia progenitors
specifically because an approximate segregation into young/prompt
and old/delayed progenitors based on SN rates has already been
observed. 

Compared to the local star formation rate (LSFR) used in 
\citet{rigault_2013} and \citet{rigault_2015},
both indicators rank galaxies similarly when the SFR is either very
high or very low. But the normalization by stellar mass provided
by LsSFR breaks an ambiguity for some intermediate cases, namely,
between lower star-formation rates in fainter regions of galaxies
and higher star-formation rates in brighter regions of galaxies.
In addition, sometimes the metric aperture in which the LSFR is
measured can extend outside the galaxy, diluting the signal. The
local stellar mass measurement is similarly diluted, thus this
dilution effect is canceled for LsSFR.

As discussed above, intrinsic SN~Ia brightnesses and colors may
well vary with progenitor age and metallicity, and the observed
brightnesses and colors are affected by dust whose properties can
covary with SN intrinsic properties. In the case of a foreground
dust screen, both the inferred SFR and stellar mass will be affected by the
similar dust. Their ratio would cancel to first order if it were not
for the fact that stellar mass measurements rely on galaxy stellar
colors to determine the $M/L$ ratio. As stellar $M/L$ is higher for
redder populations, the affect of dust is diminished for the
measurement of stellar mass. We discuss this further in
Section~\ref{sec:impact_of_dust} showing that dust extinction has no
  significant influence on our LsSFR analysis.
In addition, the LsSFR can exploit the fact that production
of dust is correlated with both higher metallicity and higher SFR
\citep[e.g.,][]{calzetti_heckman_1999}.  Since metallicity in turn
correlates with stellar mass, normalization of the SFR by stellar
mass suppresses the effect of dust on global sSFR
\citep[e.g.,][]{peek_2015}. Furthermore, 
the correlation of dust with SFR surface density
\cite[e.g.,][]{battisti_2016} suggests that any residual error in
this cancellation, while possibly leading to a distortion of the
true LsSFR as a function of observed LsSFR, will not fundamentally
change the LsSFR-based ordering of SN~Ia progenitors from
younger to older.

In the mean, the extinction-corrected sSFR inferred using H$\alpha$
in star-forming galaxies is found to have only a modest correlation
with the extinction-corrected stellar metallicity
\citep{garn_2010,salim_2014}. The relation is complex in that the
trend for galaxies with higher sSFR and lower mass opposes that in
lower sSFR and higher mass galaxies \citep{laralopez_2013}. This
leads to a rough cancellation for galaxies whose sSFR values are
typical of our SNe \citep{childress_2013b}. 

Correlation with the
amount of dust derived for each galaxy is weak
\citep[e.g.,][]{battisti_2016, peek_2015}, though this conclusion
depends somewhat on the details of how extinction of  H$\alpha$
relative to stars is handled \citep{brinchmann_2004}.  
Finally, recent
detailed integral field spectroscopical analyses of nearby galaxies
have demonstrated that the relations between extinction-corrected
stellar mass, metallicity, star formation extend down to kiloparsec
scales \citep{sanchez_2013,gonzalez_2014,navaro_2015,canodiaz_2016}.

To summarize, we expect the LsSFR indicator to probe the fraction
of young stars in the proximity of SNe~Ia, and by construction, it
should correlate only weakly with the amount of interstellar dust.
It is also expected to have only modest correlation with stellar
metallicity for most SN~Ia hosts, and if anything, opposing trends
toward higher and lower sSFRs. These properties make the LsSFR a reasonably
clean indicator of relative SN~Ia progenitor age. LsSFR has the important
added benefit of providing normalization when a metric
aperture extends outside the detectable boundaries of the host glaaxy.
Consequently, we use
LsSFR as a proxy for investigating relations between progenitor
ages and SN~Ia demographics and standardization.

\section{Measurement of LsSFR}
\label{sec:data}

All SNe, host-galaxy H$\alpha$, and some host-galaxy imaging, presented
here have been measured by the Nearby Supernova Factory (SNfactory)
using our SuperNova Integral Field Spectrograph
\cite[SNIFS;][]{aldering_2002,lantz_2004}.  Additional imaging comes
from the Sloan Digital Sky Survey.
The current SNfactory sample consists of 198 SNe~Ia
having fully-processed spectrophotometric lightcurve data, including
observations on at least two photometric nights, final references, {and a
host spectroscopic redshift. These all have at least 5 spectra while the SN is
active, and pass the quality cuts suggested by \cite{guy_2010}.  
We further limit our sample to the redshift range of $0.02<z<0.08$
needed to measure the local H$\alpha$ with SNIFS; 38 SNe~Ia are
lost due to this requirement. In addition, the $g$ and $i$
imaging of the host is required to be free of SN light so that stellar masses can be 
accurately measured; the host imaging for 13 SNe~Ia is contaminated by SN light, 
further} reducing the sample to 147 SNe~Ia.
These redshift and imaging selections are independent of SN properties. 
More than 80\% of our SNe are from searches where there was no
pre-selection based on host galaxy properties (those
whose names start with ``SNF'', ``LSQ'', or ``PTF'' in Table~\ref{tab:data}).
In addition, we exclude six SNe~Ia in the
SN~1991T, SN~1991bg and SN~2002cx subclasses, as these are considered
too peculiar, and not central to the question of environmental
effects for normal SNe~Ia.
The impact of this last cut on our result is tested
in Section~\ref{sec:lssfr_3rd_param}.  
The final sample of 141 SNe~Ia is roughly twice the size of
that used in \citet{rigault_2013}.

In the following Sections we detail the measurement of spectroscopic
and photometric data. The supernova data used in this paper correspond
to those presented in \cite{saunders2018} and \cite{leget2020} ; see
\cite{aldering2020}.
Data are corrected for Milky Way 
SN and host coordinates as well as photometric
measurements are given in
Table~\ref{tab:lowleveldata}. 

All of the derived quantities used for this analysis are given in
Table~\ref{tab:data}. 
In addition,
  online \footnote{http://snfactory.lbl.gov/snf/data}
we provide the H$\alpha$\ spectrum of the host galaxy within a 1~kpc
aperture, the samples from the posteriors used for the H$\alpha$
and stellar mass measurements, along with summary plots\footnote{These
plots show the H$\alpha$ fits as in Fig.~\ref{fig:hafit}, the
stellar mass fits as in Fig.~\ref{fig:mass_measurement} and the
LsSFR samples as in Fig.~\ref{fig:lssfr_samplers} for all SNe.}.

The lightcurve parameters were derived
using the \salt{} fitter \citep{guy_2007, betoule_2014}.  We used
relative distances determined from the redshifts to convert
SN~Ia fluxes to relative luminosities.  Host-galaxy  
redshifts come from \cite{childress_2013}, the measurement of H$\alpha$
wavelengths presented here, or the literature. These redshifts are accurate to
better than $\sigma_z = 0.001$.  

\begin{sidewaystable*}
\centering
\caption{Coordinates and measured photometric data for the \nSNe\ SNe~Ia used in this analysis. 
The complete table is available in the electronic edition.
}
\label{tab:lowleveldata}
\begin{tabular}{l|ccccccccccc}
SN Name & SN RA & SN Dec & host RA & host Dec & Local $g$ & Local $i$ &
                                                                  Global
                                                                        $g$
  & Global $i$ & Note\\
                & $\mathrm{degree}$ &$\mathrm{degree}$ &
                                                         $\mathrm{degree}$& $\mathrm{degree}$ &
$\mathrm{mag}$& $\mathrm{mag}$ & $\mathrm{mag}$ & $\mathrm{mag}$ &\\[0.15em]
\hline\\[-0.8em]
SNF20060511-014 &+315.1390&-24.6175&+315.1336&-24.6173&$23.26^{+0.44}_{-0.31}$ &$22.24^{+0.39}_{-0.29}$ &$16.45^{+0.01}_{-0.01}$ &$15.57^{+0.01}_{-0.01}$ &a \\[0.50em]
SNF20060512-001 &+213.3541&+17.7833&+213.3538&+17.7830&$19.23^{+0.01}_{-0.01}$ &$18.53^{+0.01}_{-0.01}$ &$17.44^{+0.01}_{-0.01}$ &$16.84^{+0.01}_{-0.01}$ & \\[0.50em]
SNF20060512-002 &+218.3799&+17.0200&+218.3799&+17.0199&$19.01^{+0.01}_{-0.01}$ &$17.63^{+0.01}_{-0.01}$ &$15.98^{+0.01}_{-0.01}$ &$14.86^{+0.01}_{-0.01}$ & \\[0.50em]
SNF20060521-001 &+184.2208&-3.2581&+184.2208&-3.2580&$20.37^{+0.02}_{-0.02}$ &$19.26^{+0.01}_{-0.01}$ &$18.89^{+0.01}_{-0.01}$ &$17.72^{+0.01}_{-0.01}$ & \\[0.50em]
SNF20060521-008 &+191.0160&-5.1028&+191.0125&-5.1095&$23.86^{+0.36}_{-0.27}$ &$22.94^{+0.41}_{-0.29}$ &$15.92^{+0.01}_{-0.01}$ &$14.61^{+0.01}_{-0.01}$ & \\[0.50em]
SNF20060526-003 &+220.7628&-18.8790&+220.7606&-18.8772&$24.11^{+0.39}_{-0.30}$ &$23.79^{+0.90}_{-0.51}$ &$16.08^{+0.01}_{-0.01}$ &$15.47^{+0.01}_{-0.01}$ &a \\[0.50em]
\end{tabular}
\tablefoot{ An ``a'' indicates cases where SNIFS host galaxy imaging is used. All coordinates are for equinox J2000.}
\end{sidewaystable*}

\begin{sidewaystable*}
\centering
\caption{Data for the \nSNe\ SNe~Ia used in this analysis. The complete
table is  available in the electronic edition.
}
\label{tab:data}
\begin{tabular}{l|ccccccccccc}

SN Name & $\Delta M_B^{corr}$ &$z_{CMB}$ & $c$ & $x_1$ & Local Mass & Local SFR &
Local sSFR & Global Mass & \PY & \PM & Refs\\
                & $\mathrm{mag}$ & & & &$\log(\mathrm{M_*/M_\odot})$ &
$\log(\mathrm{M_*\, yr^{-1}\,kpc^{-2}}) $ &
                                            $\log(\mathrm{yr^{-1}\,kpc^{-2}})$&
                                                                                $\log(\mathrm{M_*/M_\odot})$
                         & \% & \% & \\[0.15em]
\hline\\[-0.8em]
SNF20060511-014 & $-0.07\pm{0.14}$ & $0.0467$ & $-0.03\pm{0.03}$ & $-0.65\pm{0.17}$ & $7.55^{+0.25}_{-0.27}$ &$-3.20^{+0.10}_{-0.13}$&$-10.8^{+0.3}_{-0.3}$& $10.20^{+0.10}_{-0.10}$ & 59 & 97 & -- \\[0.50em]
SNF20060512-001 & $-0.06\pm{0.13}$ & $0.0389$ & $+0.02\pm{0.03}$ & $+0.71\pm{0.13}$ & $8.73^{+0.10}_{-0.10}$ &$-1.68^{+0.01}_{-0.01}$&$-10.4^{+0.1}_{-0.1}$& $9.32^{+0.10}_{-0.10}$ & 100 & 0 & -- \\[0.50em]
SNF20060512-002 & $-0.45\pm{0.12}$ & $0.0509$ & $+0.05\pm{0.03}$ & $-0.92\pm{0.19}$ & $9.80^{+0.10}_{-0.10}$ &$-2.00^{+0.01}_{-0.01}$&$-11.8^{+0.1}_{-0.1}$& $10.72^{+0.10}_{-0.10}$ & 0 & 100 & -- \\[0.50em]
SNF20060521-001 & $-0.03\pm{0.13}$ & $0.0679$ & $-0.05\pm{0.03}$ & $-1.15\pm{0.23}$ & $9.22^{+0.10}_{-0.10}$ &$-4.14^{+0.44}_{-0.77}$&$-13.4^{+0.5}_{-0.7}$& $9.88^{+0.10}_{-0.10}$ & 0 & 10 & -- \\[0.50em]
SNF20060521-008 & $-0.27\pm{0.14}$ & $0.0562$ & $+0.09\pm{0.03}$ & $-1.36\pm{0.20}$ & $7.41^{+0.25}_{-0.27}$ &$-3.49^{+0.13}_{-0.22}$&$-10.9^{+0.3}_{-0.4}$& $11.05^{+0.10}_{-0.10}$ & 37 & 100 & -- \\[0.50em]
SNF20060526-003 & $+0.01\pm{0.12}$ & $0.0788$ & $-0.03\pm{0.03}$ & $+0.30\pm{0.14}$ & $7.29^{+0.33}_{-0.42}$ &$-3.08^{+0.07}_{-0.09}$&$-10.4^{+0.4}_{-0.3}$& $10.52^{+0.10}_{-0.10}$ & 90 & 100 & -- \\[0.50em]

\end{tabular}
\tablefoot{ 
\\ $\Delta M_B^{corr}$ includes a $0.094\,\mathrm{mag}$ residual dispersion. 
\\ Heliocentric redshifts are taken from \cite{childress_2013b}, except for LSQ12ekl, LSQ13vy, PTF10wnm
    PTF10xyt, PTF10zdk, and PTF11mty,
\\  for which heliocentric redshifts were
    derived from our local H$\alpha$ measurements.
\\ \PY\ is the statistical classification 
    for a SN~Ia to be young, as given Eq.~\ref{eq:p(delayed)}. 
\\ \PM\ is the probability for the global stellar mass of the SN host galaxy
   to be higher than $\log(M_*/M_\odot)=10$.  
\\ The Refs column provides references for external discoveries (
   see the electronic edition).
}
\end{sidewaystable*}

\subsection{Host-galaxy identification}
\label{sec:host_id}

While the location of the aperture to measure local SFR and local
stellar mass is given by the SN, a redshift is needed in order to
define the angular size of a consistent metric aperture. In addition,
selection of the correct host galaxy is needed for measuring the
global stellar mass.  Therefore, before discussing measurement
details, we briefly describe how we associated our SNe~Ia with their
respective host galaxies.

Many of our SNe~Ia -- 122 of \nSNe\ --  have had their host galaxy
determined in \cite{childress_2013b}. As it was necessary to develop a procedure
to find hosts for the remainder, we also applied the technique to
our previously-published host identifications. Unlike in
\cite{childress_2013b}, where additional imaging was obtained when
needed, for the additional SNe we only have host data within the
SDSS footprint at this time. Therefore, we conduct our host galaxy
identification using the SDSS catalog ``V/139'' (Vizier). We search
this catalog for galaxies projected within 40~kpc of each SN, and
then define a candidate host to be the nearest galaxy based on its
effective elliptical distance \citep{gupta_2016}, as
determined by the semi-major and 
semi-minor axes -- second moments of the ellipse $a$ and $b$,
respectively -- extracted using
\sep{}\footnote{http://github.com/kbarbary/sep} \citep{barbary_2016}, the Python
implementation of \pkg{Sextractor} \citep{bertin_1996}.
If the SN is within $3\times$ the elliptical radius, defined as
$2.5\times$ the second moment ellipse, of the candidate host and the
redshifts of the candidate and SN agree, the candidate is deemed to
be the true host galaxy.

Three cases could be considered ambiguous for some applications:
\SNNGC, \SNPGC\ and SNF20080913-031. For these, a massive elliptical
galaxy is found with a redshift consistent with the SN, but outside
our initial matching distance of 40~kpc or beyond $3\times$ the
elliptical radius. Visual examination of WISE W1 images shows that
the old stellar light of these ellipticals extends out to the
SN position at very faint levels.  For this analysis we paired these
galaxies and SNe, but the reader might want to use these pairings
carefully for other applications.

\cite{childress_2013} paired SNF20070817-003 with a galaxy $\sim50\,$kpc
away whose major axis points towards the SN. However, in the course
of this study we identified a small galaxy immediately adjacent to the
SN. We do not yet have a redshift for this galaxy, so we do not use
SNF20070817-003 for the analysis here.

\subsection{Measuring the local star formation rate}
\label{sec:data_sfr}

As in \cite{rigault_2013} we measure the local star formation
rate using H$\alpha$ emission obtained using SNIFS.
SNIFS is a fully integrated instrument optimized for semi-automated
observations of SNe on the structured background typical of galaxies.
It covers the full optical window at moderate spectral resolution,
and has been continuously mounted on the University of Hawaii 2.2~m
telescope on Mauna Kea since 2004.  The integral field spectrograph
has a fully filled $6\farcs4 \times 6\farcs4$ spectroscopic
field-of-view subdivided into a grid of $15 \times 15$ contiguous
square spatial elements (spaxels).  The dual-channel spectrograph
simultaneously covers 3200--5200~\AA{} ($B$-channel) and
5100--10\,000~\AA{} ($R$-channel) with 6.6 and 7.5~\AA{} FWHM resolution,
respectively.  The method of data reduction of the $x,y,\lambda$
data cubes was summarized by \cite{aldering_2006} and updated in
Section~2.1 of \cite{scalzo_2010}.  The flux calibration methodology is described
in Section~2.2 of \cite{pereira_2013}, based on the technique for measuring
atmospheric extinction developed in \cite{buton_2013}.  For measurement
of the SN spectrophotometry, the host galaxy is subtracted as
described in \cite{bongard_2011}. SNIFS has a parallel imaging
channel equipped with $ugriz$ filters, which has been used to obtain some
additional host-galaxy imaging,
as described in \cite{childress_2013}.

The extraction of the host-galaxy local spectra from the SNfactory data
follows the recipe provided in \cite{rigault_2013}, summarized here.
The SNfactory has typically taken $\sim15$ spectra per SN~Ia, including
two ``final references'' taken at the SN location long after the
SN has faded away.
Here, we only use the final references to extract the local
host spectrum of each SN~Ia.  In \cite{rigault_2013} we also included
data cubes taken when the SN was present but had been subtracted
based on 3D PSF modeling \citep{bongard_2011,buton_2013}. These
additional data do not significantly improve the signal-to-noise
ratio of the host-galaxy data since the photon noise from the SN
dominates in most cases. We consequently set aside those data cubes
in order to avoid potential, though rare, SN contamination
of the host-galaxy measurements.

As detailed in \cite{rigault_2013}, we build a sky model from
principal component analysis of thousands of sky spectra extracted
from our standard star observations. The model is then fit to the
average spectrum of the five faintest spaxels of the SNIFS final
reference cubes since they have the least host-galaxy signal.  The
fitted sky model is then removed from the entire cube assuming a
uniform sky over the SNIFS $6\farcs4 \times 6\farcs4$\ field-of-view.
In a second pass, we extract the spectrum in a 1~kpc 
radius\footnote{The choice of a 1~kpc aperture was made in
\citet{rigault_2013} in order to maximize the size of the SN sample
given the joint constraints of the field of view of SNIFS and the
typical seeing \citep[see details in][]{rigault_2013}.} around the
SN location, taking atmospheric differential refraction into account.
The host-galaxy local spectra are then optimally combined for each
SN.

As in \cite{rigault_2013}, we use a customized version of the
University of Lyon Spectroscopic analysis
Software\footnote{http://ulyss.univ-lyon1.fr/}
\citep[\texttt{ULySS},][]{koleva_2008, koleva_2009} galaxy spectral
energy fitter to measure and remove the stellar continuum. The
H$\alpha$ and [\ion{N}{II}]${\,\lambda\lambda6548,6584}$ complex
of lines, as well as [\ion{O}{II}]${\,\lambda\lambda3726,3728}$ and
[\ion{S}{II}]${\,\lambda\lambda6716,6731}$, are then simultaneously
fit. The line profile model is Gaussian, with centers fit to a
common redshift and with all lines sharing a common width.  The
H$\beta$ and [\ion{O}{III}]${\,\lambda\lambda4959,5007}$ lines could
not be consistently fit across the sample because they often lie near
the spectrograph dichroic cross-over wavelength region.

Later in this analysis it will prove useful to have the full
posterior distributions of each fitted parameter. Therefore the
fits were performed using Markov Chain Monte
Carlo (MCMC) using the python package \pkg{emcee}
\citep[v.2.1.0,][]{foreman_2013}\footnote{https://github.com/dfm/emcee}. MCMC
requires that priors be specified; 
we use the following priors: line amplitudes, flat and positive :
redshift: Gaussian distribution centered at the best available 
redshift estimate, with a standard deviation of $\mathrm{500\,km\,s^{-1}}$
added in quadrature to the redshift uncertainties; line dispersion:
Gaussian distribution centered at the $170\,\mathrm{km\,s^{-1}}$ instrumental resolution, with a width
of $15\,\mathrm{km\,s^{-1}}$. This latter prior was trained on high
signal-to-noise data and is needed when fitting noisy data
in order to prevent fits from converging on random fluctuations or
non-physical artifacts such as sky-subtraction
residuals.

\begin{figure}
  \centering
  \includegraphics[width=\linewidth]{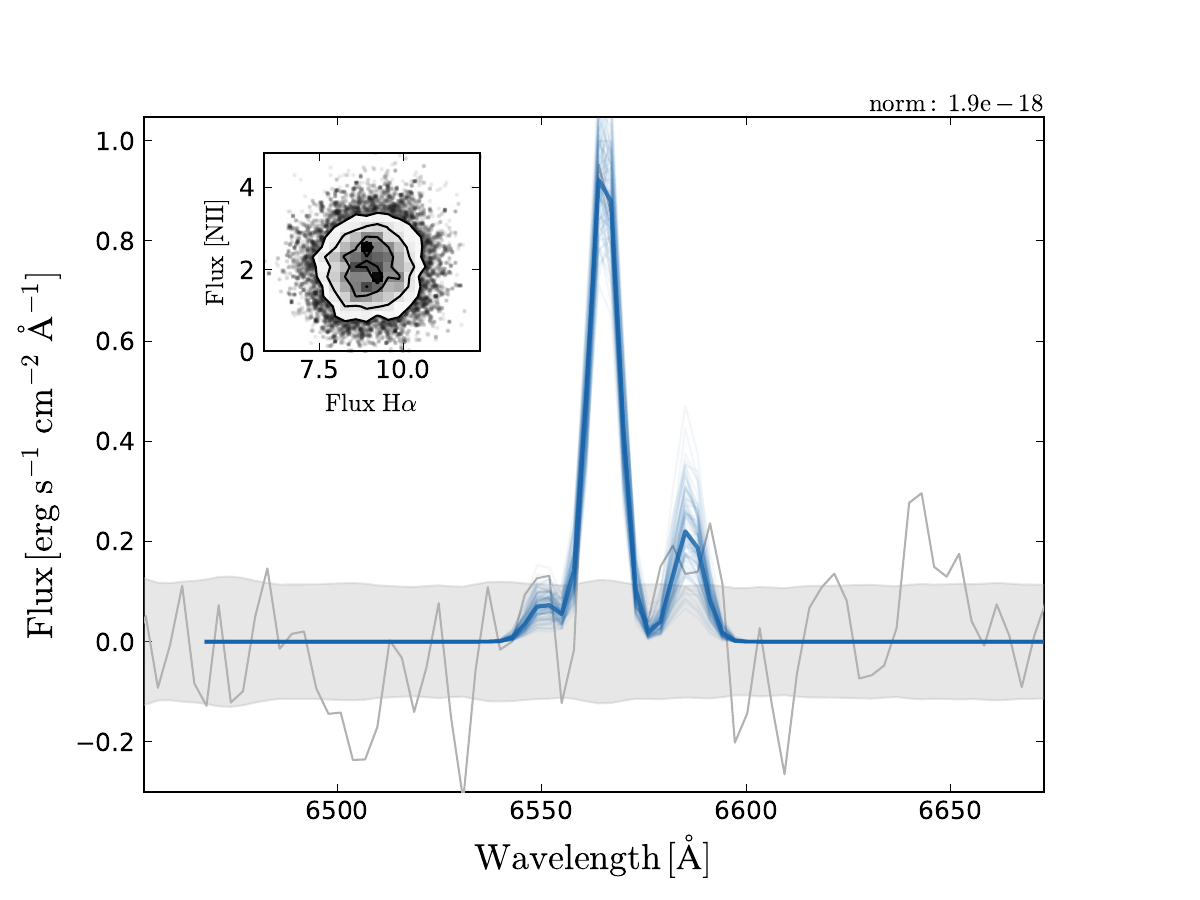}
  \caption{Illustration of a fit to the
      H$\alpha + {\rm [\ion{N}{II}]}{\,\lambda\lambda6548,6584}$ emission
      line complex for the host of SNF20060912-004, a
      typical moderate signal-to-noise case. 
      The gray line shows the emission line
      spectrum of the local host. The grey band represents its uncertainty,
      centered around zero. The thick blue line shows the best posterior
      estimation. The thin blue lines represent 100 realizations 
      from the posterior distribution, illustrating the fit
      uncertainties. A posterior density sampling in the H$\alpha$
      versus [\ion{N}{II}]  flux plane is displayed as an inset.}
  \label{fig:hafit}
\end{figure}

We use three walkers per free parameter, which we let run for
3000 iterations. We use the first 1000 iterations to burn in the
chain. Consequently each posterior distribution has 48000 samples.
Visual inspection confirmed that all fits had converged. A
typical fit of the H$\alpha + {\rm [\ion{N}{II}]}{\,\lambda\lambda6548,6584}$
emission line complex is shown in Fig.~\ref{fig:hafit}. 
The posterior distributions of the H$\alpha$ fits for the
host galaxies of the \nSNe\ SNe~Ia are available
online\footnote{http://snfactory.lbl.gov/snf/data}. 

The H$\alpha$ measurements provided here are in units of luminosity
($\mathrm{erg/s}$) per kpc$^2$.\footnote{In the course of this
analysis we discovered that the line measurements in \cite{rigault_2013}
were measured as the surface brightnesses (in $\mathrm{arcsec^{-2}}$)
averaged over a 1~kpc radius aperture, rather than the intended
luminosity per kpc$^2$.  The derived values are similar since
$1\,\mathrm{kpc} \sim 1\,\mathrm{arcsec}$ at our median redshift
$z=0.05$ ; see also erratum of \cite{rigault_2013}} As in \cite{rigault_2015},
the resulting H$\alpha$ luminosity is converted into a star formation
rate (SFR) using the \cite{calzetti_2013} calibration:
\begin{equation}
 \mathrm{SFR}(\mathrm{H}\alpha) = 5.45 \times 10^{-42}\ 
 L(\mathrm{H}\alpha)\, \mathrm{[erg\ s^{-1}]}.
\end{equation}
Since we use the full posterior distribution for each fit,
every H$\alpha$ MCMC sample is converted into a SFR sample in this
way.

This conversion assumes that H$\alpha$ is due to \ion{H}{II}
regions and not AGN emission. 
The Baldwin-Phillips-Terlevich
\citep[][BPT]{baldwin_1981} diagram can be used to distinguish
between these cases based on the [\ion{O}{III}]/H$\beta$ and
[\ion{N}{II}]/H$\alpha$ spectral line ratios.  [\ion{O}{III}]/H$\beta$
is not available, but most of the classification constraint comes
from [\ion{N}{II}]/H$\alpha$ \citep[see, e.g., Fig. 1 of ][]{kewley_2006}.
An  AGN classification is justified if the flux ratio
$\log(\mathrm{[\ion{N}{II}]\,\lambda6548}/\mathrm{(H\alpha)}) >
-0.1$.
We measure this flux ratio by computing the fraction
of $\mathrm{[\ion{N}{II}]\,\lambda6548}$ and $\mathrm{H\alpha}$ MCMC samples
that have a flux ratio greater than $-0.1\,\mathrm{dex}$.
Next, we look for cases where LsSFR might be contaminated by AGN
emission. Since we can not know what fraction of the observed
H$\alpha$ signal should be assigned to star formation, such cases could
later affect the age classification 
(Section~\ref{sec:categorizing_measurement}).
Among the initial SN sample, we identified two cases where light from the galaxy
center is contained within the 1~kpc aperture and where
[\ion{N}{II}]/H$\alpha$ indicates a possible AGN. These are SN2006ob
and SNF20060512-002. \cite{childress_2013b} obtained long-slit
spectra covering the cores of SN2006ob and SNF20060512-002, finding
[\ion{N}{II}]/H$\alpha$ [\ion{O}{III}]/H$\beta$ values indicative
of AGN activity. However, in these two cases the H$\alpha$ contained
within our aperture is too weak to pass the threshold established
later for a young system even if the H$\alpha$ were entirely from
star formation. Therefore, we retain them, 
resulting in no SNe~Ia lost because of AGN contamination.

\subsection{Measuring the local and global stellar mass}
\label{sec:mass_measurement}

Stellar mass measurements from broadband imaging require a simultaneous
determination of the stellar mass-to-light ratio and the dust
extinction.  In \citet{childress_2013b} we compared how the results
for typical SN~Ia host galaxies depended on the bandpasses available.
The most reliable results use optical, UV, and NIR data, but remain
unbiased even when only $g$- and $i$-band data are used. Here we
employ stellar masses derived from $g$- and $i$-band imaging, since
this exists and has the necessary spatial resolution for the 1~kpc
local aperture that we intend to use.

For the present study we only require internal consistency, and
therefore it is not necessary to explore the effects of different
initial mass functions, dust models, or stellar libraries. This 
enables us to employ the simple relation given in Eq.~8 of \cite{taylor_2011}:
\begin{equation}
\label{eq:taylor_mass}
\log (M_*/M_\odot) = 1.15 + 0.70\,(g - i) - 0.4\,M_i ,
\end{equation}
where $M_i$ is the absolute $i$-band AB-magnitude.
Eq.~\ref{eq:taylor_mass} was constructed using Bayesian fitting of
composite stellar populations models to $ugriz$ photometry of the
GAlaxy Mass Assembly (GAMA) sample. These models employ the
\cite{bruzual_2003} synthetic stellar population library, allow
only smooth exponentially-declining star formation histories, adopt
the \cite{chabrier_2003} initial mass function, and assume the
extinction curve of \cite{calzetti_2000}.
According to \cite{taylor_2011} this relation produces an unbiased
estimate of galaxy stellar mass with a precision of $0.1\,\mathrm{dex}$.

The $g$- and $i$-band optical imaging used to derive the local and
host-galaxy global stellar masses come from SDSS \citep[DR12,][]{sdss_dr12} 
for 118 of our SNe~Ia, and from SNIFS for
another 23 of them \citep{childress_2013b}. \footnote{SNIFS Images are
  available at http://snfactory.lbl.gov/snf/data.}
Of the SNIFS images, 20 are used for SNe~Ia that fall outside the SDSS footprint.
In addition, we cannot use the SDSS data for seven cases in which
the SDSS images were taken between $-18$ days and $+365$~days relative
to the SN  peak in B-band.  For three of these we have SNIFS imaging
data taken long after the SN faded, so were able to retain them.

The SDSS images as provided are background subtracted with calibrated
flux and astrometry.  Consequently we do not subtract any additional
background or perform further rectification.  The uncertainty images
are reconstructed following the recipe provided by the
SDSS-collaboration\footnote{http://data.sdss3.org/datamodel/files/\\
BOSS\_PHOTOOBJ/frames/RERUN/RUN/CAMCOL/frame.html}. The SNIFS images
and their associated uncertainties already exist from
\cite{childress_2013b}.  These images have calibrated
fluxes and astrometry, as detailed in Section~2.2 of
\cite{childress_2013b}.

\begin{figure}
  \centering
  \includegraphics[width=\linewidth]{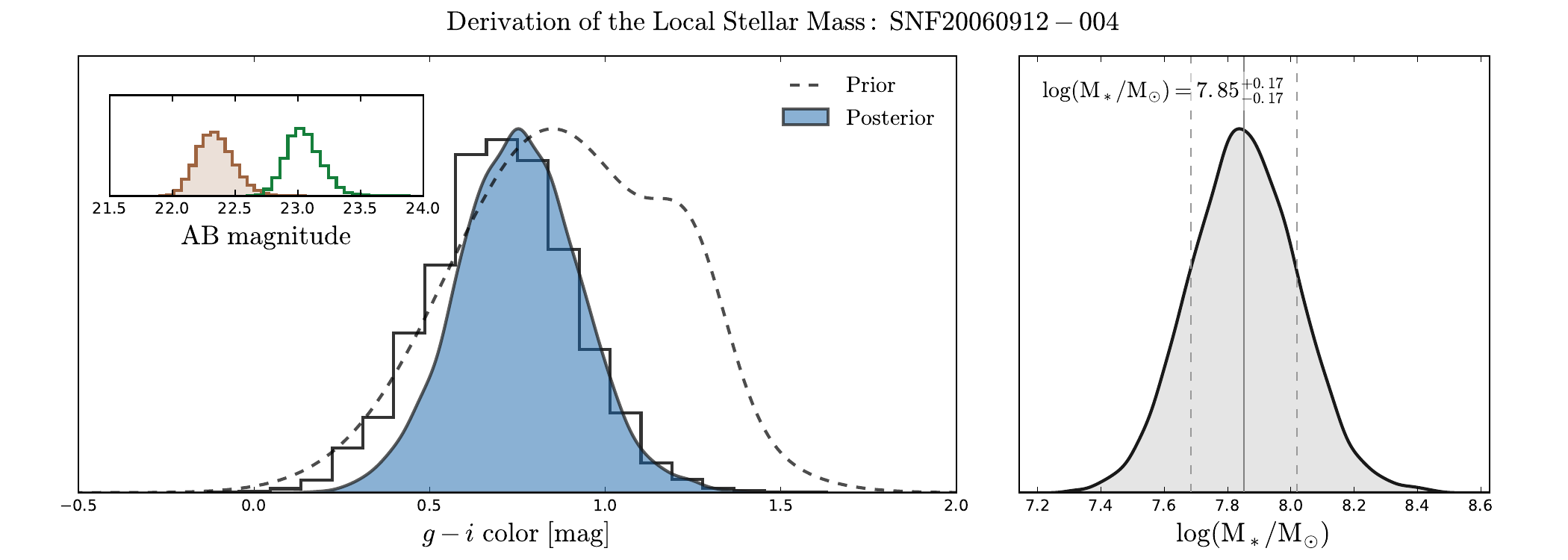}
  \caption{Illustration of how local stellar mass is derived, for the case of a
    typical moderate signal-to-noise case -- the host of SNF20060912-004.
  \emph{Left:} $(g-i)$ color distributions: the histogram shows the 
  likelihood distribution measured from the individual $g$ and $i$
  magnitude distributions shown in the inset plot (where the open green
  histogram represents $g$ and the filled brown histogram represents
  $i$). 
  The dashed line shows the prior distribution and the filled
  blue envelope shows the reconstruction of the $(g-i)$ posterior distribution.
  \emph{Right}: The posterior distribution of the local stellar
  mass; the vertical grey solid line indicates the median of 
  the distribution and the two dashed lines show the 16th and 84th percentile
  values.}
  \label{fig:mass_measurement}
\end{figure}

Eq.~\ref{eq:taylor_mass} requires a $(g-i)$ color, whose 
uncertainties are non-Gaussian due to the transformation of 
Gaussian flux uncertainties to magnitudes. In addition,
some local stellar mass measurements are in regions of lower surface
brightness that can be noisy. Therefore, we employ an informative prior for the
$(g-i)$ color distribution.  We constructed this prior using the $(g-i)$
colors of well-measured host galaxies, that is, those with a color likelihood
distribution having an RMS in $(g-i)$ less than $0.1\,\mathrm{mag}$.
This prior is illustrated in Fig.~\ref{fig:mass_measurement}.  
We tested the stability of our mass measurements and
  associated results against the manner in which the prior was built 
by alternatively using a flat prior
ranging between $-0.5<(g-i)<2$, a Gaussian prior centered on
$(g-i)=0.7\,\mathrm{mag}$ having $\mathrm{FWHM} = 0.5\,\mathrm{mag}$,
as well as a bluer prior derived from field
galaxies from \citet[][see their Fig.~4]{lange_2015}. 
We find that the local stellar masses derived using these 
different priors are consistent within a few percent of the stellar
mass error bars.

We measure the local stellar mass in the projected
1~kpc radius circular aperture centered on the SN location.
The first step is to measure the $g$ and $i$ fluxes and determine
the uncertainties within this
circular aperture, for which we use the \textsc{sum\_circle} method of 
\sep{}.
Unlike the case for faint galaxies observed in the UV with
{\it GALEX} by \citet{rigault_2015} and \citet{jones_2015}, where
use of a Poisson error model was essential due to the low numbers
of counts, for the optical observations used here the combination
of comparatively brighter sky and larger detector noise produce a
symmetric Poisson distribution that is consistent with a Gaussian.
The probability distribution for the flux measurement in a given
band can therefore be characterized by a mean corresponding to the
number of photo-electrons from the host or host region after sky
subtraction and a standard deviation set by the square-root of the
quadrature sum of the number of photo-electrons from the host, or
host region, and the sky, and variance from the detector.

The probability distribution on the mass measurement are
  non-Gaussian due to the conversion between flux and magnitudes 
in Eq.~\ref{eq:taylor_mass}, as well as non-analytic due to our use
of a prior. Therefore, we construct the posterior distribution of
the stellar mass for each individual SN using a conventional Gibbs
sampling method, which is based on Monte Carlo draws from the
measurement probability distribution functions and the prior. First,
we randomly draw $N=5000$ samples each from $g$ and $i$ flux
Gaussian probability distribution functions.  Each of these samples
is then converted to AB magnitude using either the zeropoint
calibration of $22.5\,\mathrm{mag}$ provided by SDSS or the zeropoint
calibration provided by \cite{childress_2013b} for SNIFS.  An example
of the resulting $g$ and $i$ magnitude distributions is shown for
a typical SN host galaxy in the inset of Fig.~\ref{fig:mass_measurement}.
Samples from these distribution are combined to obtain the $(g-i)$
likelihood function. This likelihood function is combined with the
$(g-i)$ prior to obtain the $(g-i)$ posterior distribution.  To
construct stellar masses we combine samples from the $i$ magnitude
distribution with an equal number of samples from the $(g-i)$
posterior distribution.  For these steps, we use a kernal density
estimator to sample from the $(g-i)$ posterior distribution.  We
then apply Eq.~\ref{eq:taylor_mass} to obtain $N$ stellar mass
samples for each SN host galaxy.  To these we add random Gaussian
noise of $0.1\,\mathrm{dex}$ to account for the scatter in
Eq.~\ref{eq:taylor_mass} found by \cite{taylor_2011} for the GAMA
sample.  This calibration noise dominates the measurement uncertainties
on host-galaxy global stellar masses (see Table~\ref{tab:data}).
Each stellar mass reported in Table~\ref{tab:data} is
then the mean of this posterior distribution, and the reported
uncertainties are the $1\,\sigma$ (16th and 84th percentiles) of
this posterior. The entire stellar mass derivation process is
illustrated in Fig.~\ref{fig:mass_measurement}, and was settled
before Hubble residuals were examined.

For host-galaxy global stellar mass measurements we use the integrated magnitudes
from the SDSS catalog.  We tested the consistency of our mass
derivation procedure by comparing with global stellar masses from
\cite{childress_2013} for the SNe~Ia common to both samples.  Our
measurements are compatible: the mean of the $\Delta \log(M_*/M_\odot)$
pull distribution is compatible with zero ($0.034\pm0.022$) and 
its standard deviation is compatible with unity ($1.17 \pm 0.08$). 
The local and global stellar masses are given in Table~\ref{tab:data},
in units of $\log(M_*/M_\odot)$.

\subsection{Categorizing SNe~Ia by age}
\label{sec:categorizing_measurement}

LsSFR is designed to estimate the fraction of young versus old stars
projected onto the host-galaxy region in the vicinity of the SN
location. 
Hence, based upon the converging evidence that the SN~Ia observed
progenitor age distribution is bimodal (cf. discussion in the
Introduction) , we use the LsSFR as a running
variable to categorize each SN~Ia as being younger or older.

As discussed in Section~\ref{sec:introduction} of this paper 
and the Introduction of \cite{rigault_2015},
the local approach is especially relevant in this context since a
young progenitor do not have time to disperse far from the
environment from which it originates. For instance, assuming
the worst case of pure linear expansion, stars in a birth cluster
require $\sim300$~Myr to dispersion by 1~kpc given their typical
$\sim3$~km~s$^{-2}$ initial velocity dispersion. In practice, in
rotationally-supported galaxies much of this motion is epicyclic
within the disk, thus extending the dispersal time. The recent
study by \citet{aramyan_2016} finds that $\sim 66$\% of SNe~Ia in
spirals are associated with spiral arms, where most star-formation
takes place. Given that in the local universe roughly 60\% of SNe~Ia
occur in spirals \citep{li_2011}, this implies that $\sim 40$\% of
all SNe~Ia are associated with spiral arms. In addition, normalization
by the relative amount of stars -- the local stellar mass in the
denominator of LsSFR -- accounts for the probability that an older
SN~Ia is projected onto, or has wandered into, a region of star
formation.

\begin{figure}
  \includegraphics[width=\linewidth]{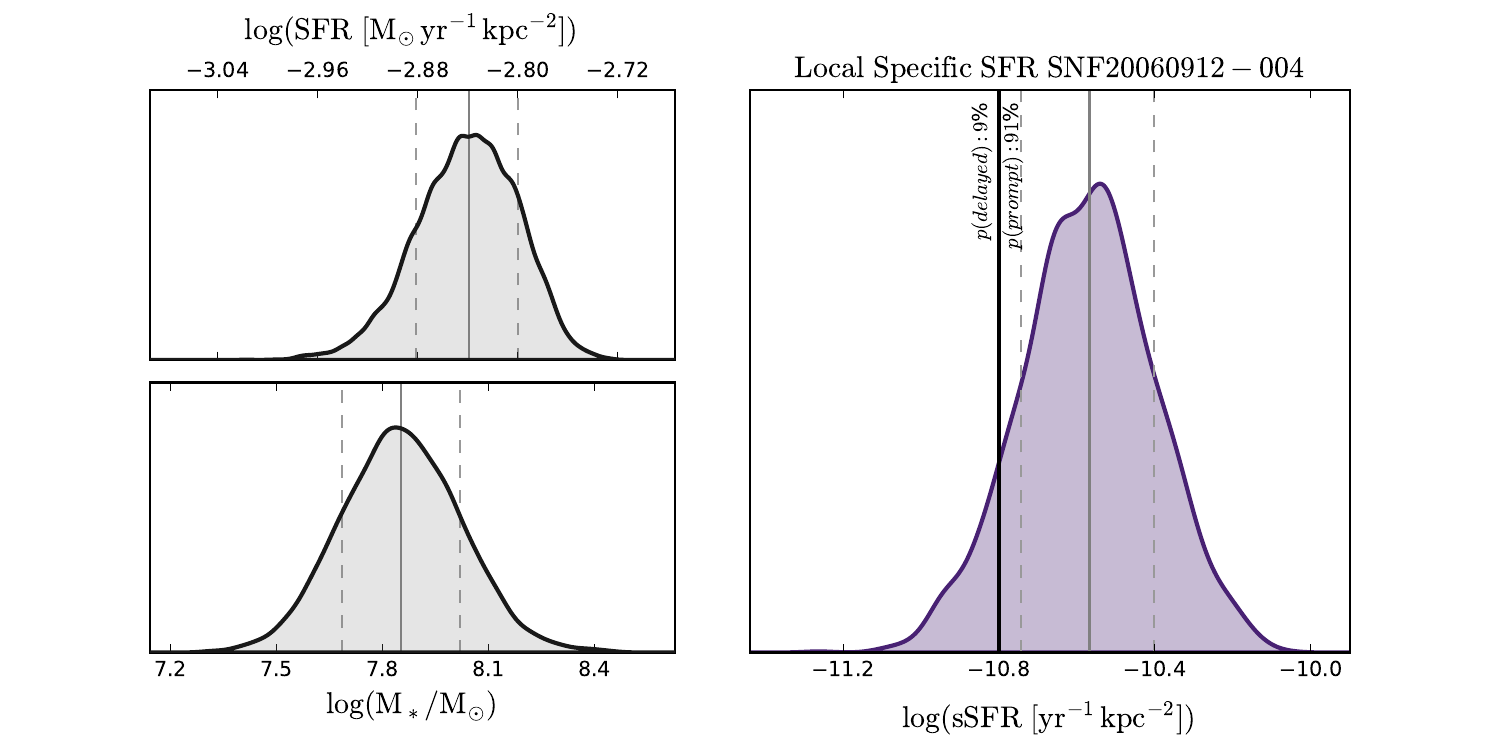}
  \caption{Illustration of how measurement uncertainties taken from posterior
distributions for local SFR and local stellar mass are used to construct
the posterior distribution for LsSFR.
\emph{Top left:} the local $\log(\mathrm{SFR})$; 
    \emph{Lower left:} the local
    $\log(\mathrm{M_*/M_\odot})$ (see also
    Fig.~\ref{fig:mass_measurement}); 
    \emph{Right:} the resulting local $\log(\mathrm{sSFR})$. 
    The vertical grey solid lines indicate the median of
    each distribution, and the two dashed lines delimit the 16th to 84th percentile
    range. On the LsSFR plot,  the thick vertical black line shows 
    $\log(\mathrm{LsSFR_{cut}})=-10.8$. 
    This figure again exemplifies a
    moderate signal to noise ratio case using the host galaxy for SNF20060912-004.}
  \label{fig:lssfr_samplers}
\end{figure}

To classify each SN, we divide the sample relative to a threshold.
Since the LsSFR measurements have uncertainties, we make use of the
LsSFR posterior distribution. The posterior is constructed by taking
the ratios of the $N$ local SFR samples of
Section~\ref{sec:data_sfr} and the $N$ local stellar mass
samples from Section~\ref{sec:mass_measurement}.  This process is
illustrated in Fig.~\ref{fig:lssfr_samplers}.  Accounting for
measurement errors in this way, we classify the SN~Ia as being
young as follows: 
\begin{equation}
\label{eq:p(delayed)}
\PY = \mathcal{P}(\mathrm{LsSFR>LsSFR_{cut}})
\end{equation}
where $\mathcal{P}(\mathrm{LsSFR>LsSFR_{cut}})$ is the fraction of LsSFR
samples from the posterior having a value greater than a chosen
threshold, $\mathrm{LsSFR_{cut}}$ (see Fig.~\ref{fig:lssfr_samplers}).

Following the decision made in \cite{rigault_2013}, the
value of $\mathrm{LsSFR_{cut}}$ was set such that 50\% of the
sum over all $\PY(\mathrm{LsSFR})$ is assigned to a younger population
and the other 50\% is assigned to an older population. 
This occurs for $\log(\mathrm{LsSFR})=-10.8$.  Having half of the
SNe~Ia in one mode or the other is compatible with DTD analyses
\citep{mannucci_2006,rodney_2014} for our redshifts ($0.02<z<0.08$),
and with the fraction of SNe~Ia associated with spiral arms
in their host galaxies \citep{aramyan_2016}.  We show in
Section~\ref{sec:lssfr_3rd_param} that our results do not significantly
vary if we change this division fraction over a range from 40\% to
60\%.

\section{Results}
\label{sec:results}

Here we examine SN~Ia demographics and standardization relative to
LsSFR and $\PY$.  We start by analyzing the distribution of lightcurve
parameters (Section~\ref{sec:result_lightcurve}) relative to LsSFR
and $\PY$.  We then study correlations with standardized Hubble
residuals (Section~\ref{sec:lssfr_bias}) segregated into younger
and older categories using $\PY$.  In Section~\ref{sec:mass_step},
we explore the connection between LsSFR and the step in Hubble
residuals with global stellar mass.  Finally, we test for differences
in the SN standardization between younger and older progenitor
populations in Section~\ref{sec:standardization}.

Throughout the entire analysis we treat the supernovae statistically,
apportioning them to the younger or older group based on
their $\PY$ values.  
However, some analyses require that each SN belong to a distinct category.
This is the case for Kolmogorov-Smirnov (KS) test
and for performing standardization independently for the two progenitor-age groups
in Section~\ref{sec:standardization}. For these cases, SNe~Ia 
having $\PY>50\%$ and $\PY<50\%$ are assumed to be younger
and older, respectively. 

In order to ensure that the results are not pulled by outliers, we
apply the Grubb criterion to identify potential outliers. Its advantage
over commonly-used $\sigma$-clipping is that it accounts for sample
size.  For our sample of \nSNe\ SNe~Ia, the Grubb criterion is equivalent
to $3.5\,\sigma$ for a normal distribution.  This criterion equates
with that of Chauvenet for a significance level rejection parameter
$\alpha=0.07$ \citep{rest_2014}. The Chauvenet, Grubb
and similar criteria are designed to identify only one outlier. This
does not affect our analysis since we would not have found any
additional outliers by relaxing this constraint.  Indeed, the only
analysis in which the Grubb criterion proposed an outlier is for
\salt{} standardization using only the young population in
Section~\ref{sec:standardization}.

We emphasize the lack of tuning in this study: the sample
selection was driven by external constraints (see Section~\ref{sec:data});
the aperture size for local measurements of the host galaxy was
dictated by SNIFS characteristics (see Section~\ref{sec:data_sfr})
and is the same as implemented in \cite{rigault_2013}; division of
the sample into equal halves follows the method established in
\cite{rigault_2013}; use of a step in Hubble residuals as the
underlying model follows the demonstration in \cite{childress_2013}
that a step best describes the data, and the subsequent use of a
Hubble-residual step in \cite{rigault_2013}.

\subsection{Lightcurve parameters}
\label{sec:result_lightcurve}
\begin{figure*}
  \centering
  \includegraphics[width=\linewidth]{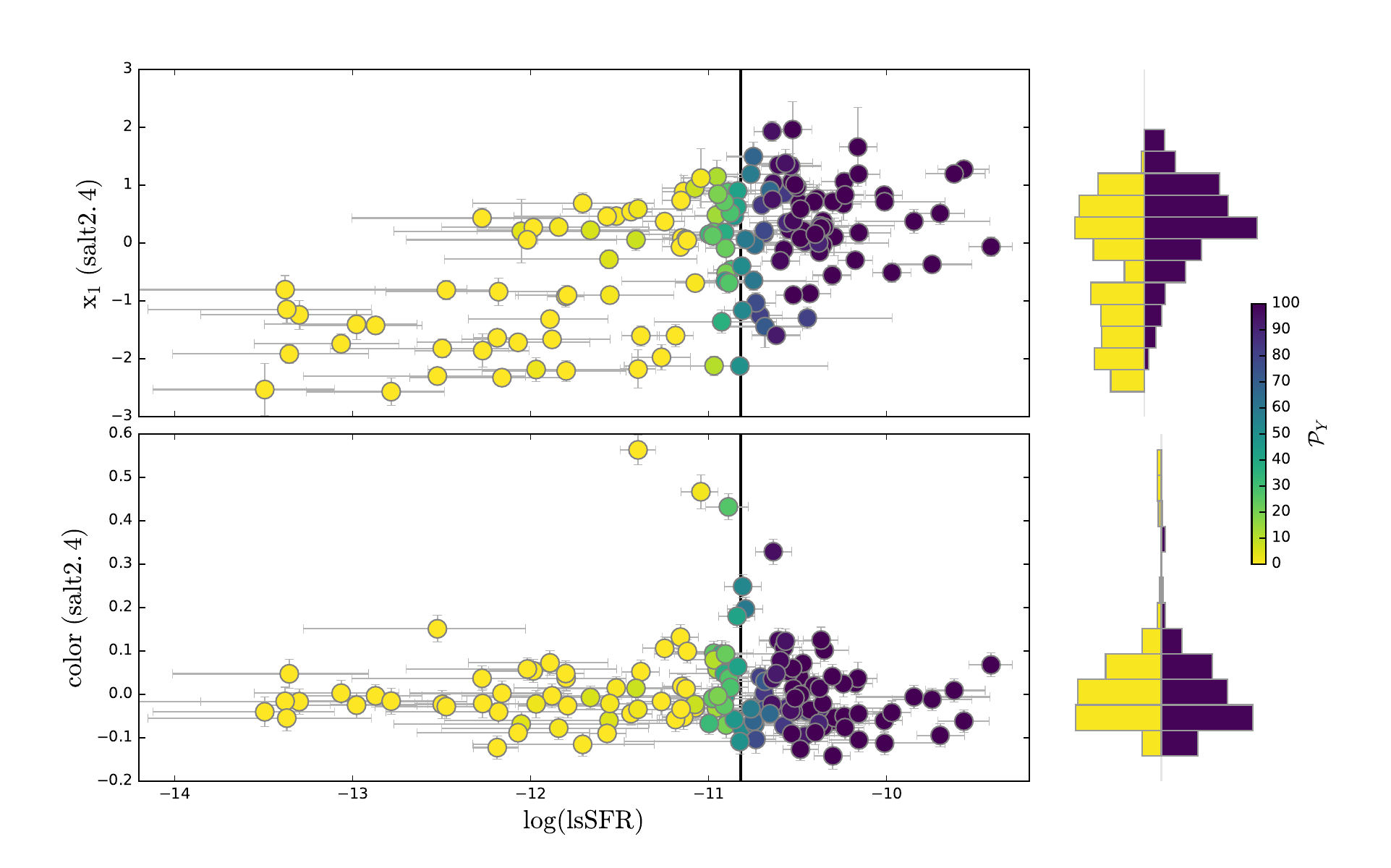}
  \caption{\salt{} $x_1$ (top) and color ``c'' (bottom) lightcurve parameters as a
    function of $\log(\mathrm{LsSFR})$. The marker-color
    represents the probability for a supernova to have a
    younger/prompt progenitor ($\PY$, see color-bar). The histograms on the right
    are $\PY$-weighted marginalization of SNe~Ia lightcurve parameters:
    toward left for the older/delayed distributions and toward right for the
    younger/prompt distributions.}
  \label{fig:lssfr_lcparam}
\end{figure*}

The distributions of the SN~Ia lightcurve parameters $x_1$ and
$c$, and LsSFR, are shown in Fig.~\ref{fig:lssfr_lcparam} and discussed here. 

\subsubsection{Lightcurve stretch}
\label{sec:result_lssfr_stretch}

Fig.~\ref{fig:lssfr_lcparam}(top) shows the \salt{} lightcurve
stretch, $x_1$, versus LsSFR.  We find that $x_1$ is correlated
with LsSFR; a Spearman rank correlation test between $x_1$ and LsSFR
gives $r_s=-0.48$, the random probability of which is less than
$10^{-9}$. This result, significant at $6.5\,\sigma$, confirms
previous findings \citep[e.g.,][]{hamuy_1996, sullivan_2010,
lampeitl_2010,rigault_2013} that the SN~Ia lightcurve stretch
distribution tracks an intrinsic SN property that depends on the
progenitor age.  While a correlation clearly exists, its scatter
is much larger than the measurement uncertainties, indicating that
other, latent, progenitor properties also are important.

In the histograms shown on the right in Fig.~\ref{fig:lssfr_lcparam},
we see that the $x_1$ dispersion is $\sim$30\% lower for the younger population
indicative of an intrinsically more homogeneous population.  The
lightcurve evolution of younger SNe~Ia is slower (greater $x_1$)
and they mainly populate the positive $x_1$ region. In contrast,
the older population seems to populate the entire $x_1$ range. A
KS-test confirms that the $x_1$ distributions are inconsistent,
giving a probability less than $10^{-4}$ that they arise
from the same parent distribution. However, after removing the
already-detected difference in the means, the shapes of the distribution
have a 7\% probability of being consistent.

There are established, but still rather qualitative, connections
between lightcurve stretch and SN~Ia progenitor channels.  When
restricted to the single-degenerate progenitor channel, where the
total ejecta mass is very nearly the Chandrasekhar mass, lightcurve
stretch is usually interpreted as a indicator of the mass of
radioactive $^{56}$Ni produced in the explosion and which subsequently
powers the lightcurve. Alternatively, reconstruction of progenitor
properties based on bolometric lightcurves and velocities in which
the total ejecta mass is not restricted indicate that lightcurve
stretch is most strongly correlated with total ejecta mass
\cite{scalzo_2014}.  The correlation with LsSFR or age could then
be connected with the subset of binary system parameters, such as
separation and relative masses, that affect the timescale for
inducing a SN~Ia.

\subsubsection{Lightcurve color}
\label{sec:result_lssfr_color}

Fig.~\ref{fig:lssfr_lcparam}(bottom) shows the \salt{} lightcurve
color, $c$ versus LsSFR.  The young/prompt SNe appear $\Delta c=0.047\pm0.017$~mag 
bluer than the old/delayed SNe, but the reddest SN, SNF20061022-014,
is primarily responsible for the offset.  Removing it reduces $\Delta
c$ to $-0.020 \pm 0.015$~mag.  The core of the color distributions, i.e., without the \nRed\ SNe with $c>0.2$, show no sign of
difference between the younger and older with $\Delta c = 0.008 \pm
0.013$~mag.  More generally, we find no significant correlation
between $c$ and LsSFR, as indicated by a Spearman rank correlation
coefficient $r_s=-0.11$, which deviates from zero by only
$\sim1.3\,\sigma$. This finding is in agreement with studies
based on global stellar host properties
\citep[e.g.,][]{sullivan_2010,lampeitl_2010,pan_2014}. The weakness
of the observed trends suggests that the progenitor age does not
have significant influence on the SN color as given by the \salt{}
lightcurve fitter. Removing the \nRed\ reddest SNe does not
change this result.

\subsection{Standardization using LsSFR}
\label{sec:lssfr_bias}

The measurement that is directly used for SN~Ia cosmology is the
standardized brightness. Systematic deviations from a best-fit
cosmology can be used to help uncover effects not fully accounted
for in the standardization process. The most commonly used
standardization uses a linear combination of the lightcurve stretch
and color \citep{tripp_1998}. More recent variants have included
the global stellar mass step, as well as non-linear relations in stretch and/or
color \citep{rubin_unity_2015,scolnic_2016}.

In this subsection, we begin by standardizing our SNe~Ia using linear
relations between the SN~Ia peak magnitudes, stretch and color
produced by \salt{}.  The residuals from the Hubble diagram are then
referred to as $\Delta M_B^{corr}$, which in the \salt{} framework are
given by:
\begin{equation}
\label{eq:standardization_2params}
  \Delta M_B^{corr} = \Delta M_B +\alpha \times x_1 - \beta \times c,
\end{equation}
where, $\Delta M_B$ is the observed difference of absolute SN
magnitudes in $B$-band, $\alpha$ and $\beta$ are the -- blinded --
standardization coefficients for stretch, $x_1$, and color, $c$,
respectively. In a second step, we also include the probability
that a supernova is young ($\PY$) as a third standardization parameter
(see Section~\ref{sec:lssfr_3rd_param}). In all of these fits,
the full matrix of measurement covariances is used. The main results of this
subsection are summarized in Table~\ref{tab:summary_standardization}.

\subsubsection{LsSFR step measurement}
\label{sec:lssfr_step}

\begin{figure*}
  \centering
  \includegraphics[width=\linewidth]{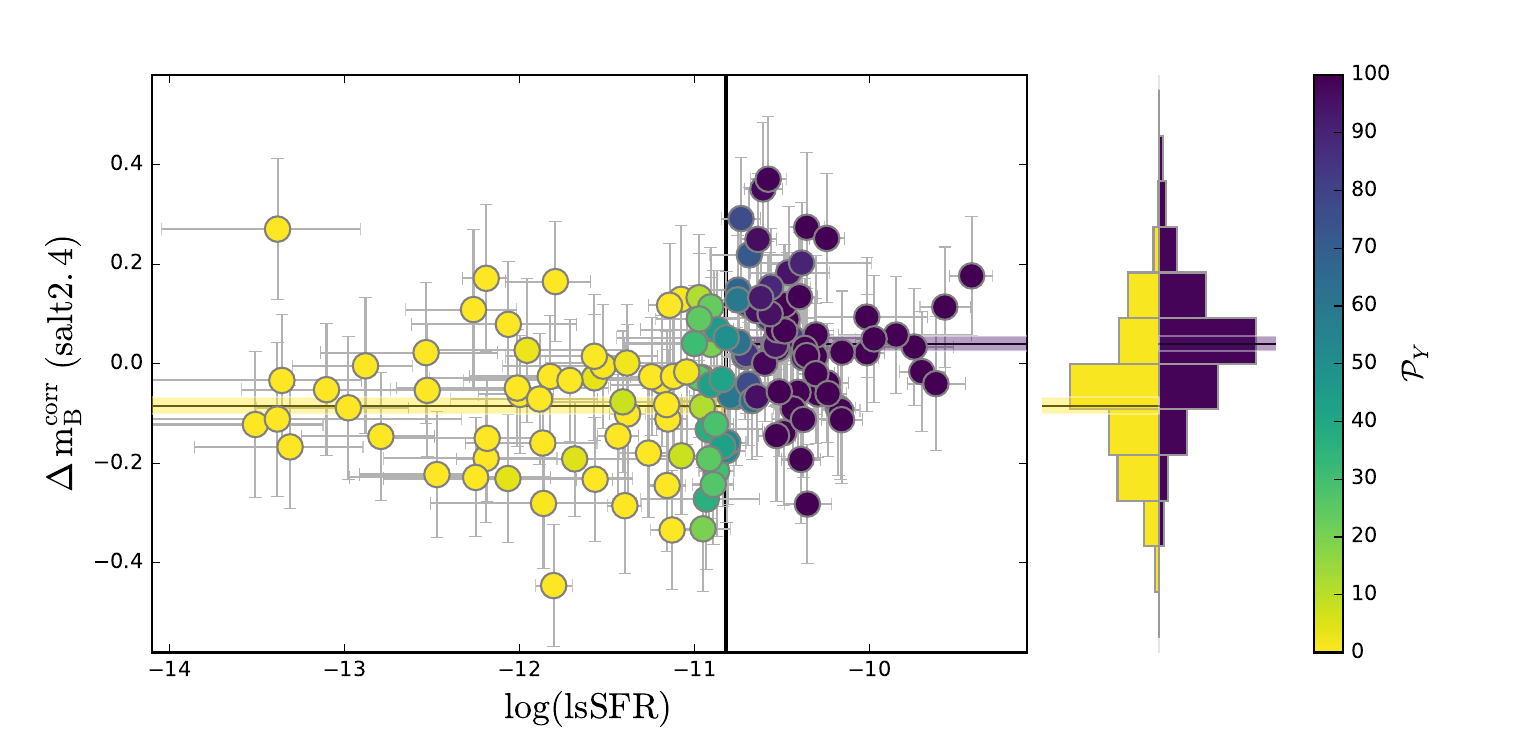}
  \caption{
    SN~Ia Hubble residuals, $\Delta M_B^{corr}$,
    as a function of $\log(\mathrm{LsSFR})$, calculated from
    a conventional linear standardization using \salt{} 
    lightcurve parameters. The plot symbols and histograms 
    follow the rules of Fig.~\ref{fig:lssfr_lcparam}. 
    In the main panel and in the histogram-panel, the two horizontal
    bands show the weighted average of $\Delta M_B^{corr}$ per 
    progenitor age group.
    The width of each band represents the corresponding
    error on the mean, and their offset illustrates
    the Hubble residual offset between the two age groups.
    The error bars on $\Delta M_B^{corr}$ include the
    measurement, \salt{} systematic, and residual dispersions
    from the maximum-likelihood fit to each population.}
  \label{fig:lssfr_step}
\end{figure*}

The correlation between $\log(\mathrm{LsSFR})$ and $\Delta M_B^{corr}$
is presented Figure~\ref{fig:lssfr_step}.  A sharp transition in
Hubble residuals is clearly visible around $\log(\mathrm{LsSFR})\sim-10.8$.
To assess the size of this step, we perform a maximum likelihood
fit for the Hubble residual step between these two populations,
modeled as two independent normal distributions each having its own
mean and standard deviation as free parameters. Details of this
procedure are given in Appendix~\ref{app:fitstep}. This fit gives
a \salt{} Hubble residual offset of $\DY=0.125\pm0.023\, \mathrm{mag}$,
in which the younger SNe~Ia are fainter. This result
is incompatible with no LsSFR step at $5.5\,\sigma$.

The histograms plotted on the right in Fig.~\ref{fig:lssfr_step}
show that the individual populations appear normally distributed,
and there is no evidence that the difference in means is pulled by
outliers. The residual dispersions, after accounting for measurement
error, are similar with $\sigma_{resid}=0.103\pm0.015\,\mathrm{mag}$
for the young subpopulation versus $\sigma_{resid}=0.115\pm0.015\,\mathrm{mag}$
for the old, including the 0.055~mag 
systematic lightcurve fitting error given by \salt{} (see details in
Appendix~ \ref{app:fitstep}). 

The difference in mean Hubble residual between the two groups is further
supported when comparing their $\Delta M_B^{corr}$ distributions.
A  KS-test finds that the probability that both
$\Delta M_B^{corr}$-distributions arise from the same underlying
distribution is $10^{-5}$.

This is the most significant detection of a standardized SN~Ia
brightness systematic connected to host-galaxy environment measured
to date. This suggests that the conceptual motivation for constructing
the LsSFR metric -- as an attempt to account for both a young progenitor
population associated with star formation and an older progenitor
population traced by stellar mass using the immediate SN environment
-- has merit.

\subsubsection{\PY\ as a third standardization parameter}
\label{sec:lssfr_3rd_param}

The fits performed so far were done sequentially in order to get a
first look at the effect due to LsSFR, in a fashion analogous to past
studies of the global stellar mass step \citep{kelly_2010, sullivan_2010,
gupta_2011,childress_2013}. The proper approach for a quantitative
result is to perform a fit for \DY\ and the lightcurve
parameter standardization coefficients simultaneously.  This is the
approach currently used when including the host-galaxy global stellar mass as a third
standardization parameter \citep[e.g.,][]{suzuki_2012,betoule_2014}.
To do this we use $\PY$ to segregate the populations. 

Since the measurement of $\PY$ is completely independent
of the SN lightcurve fits, there is no measurement covariance between
$\PY$ and the lightcurve parameters. Therefore, these off-diagonal
terms are set to zero in the covariance matrix used for our fit.
(Recall that the presence of a correlation in measurement values
does not imply covariance in the measurement uncertainties.) The
resulting LsSFR step is now $\DY=0.163\pm0.029\,\mathrm{mag}$,
which is higher than that found in Section~\ref{sec:lssfr_step}
when performing the standardization fit sequentially. The significance
increases slightly, to $5.7\,\sigma$.  Instead fitting a line as a
function of LsSFR, including errors on LsSFR, gives a slope of
$0.079\pm 0.018\,\mathrm{mag/dex}$.  The significance of the slope
is $4.3\,\sigma$. This is much less than that for the step, and
thus a step is significantly favored by the data.

\begin{table*}
\centering
\caption{Summary of the SNe~Ia Standardization. 
}
\label{tab:summary_standardization}
\begin{tabular}{l | c c c c}
\hline\hline\\[-0.8em]
\hline\\[-0.8em]
Parameters  & wRMS & $\sigma_{resid}$ & \DM & \DY \\[0.15em]
\hline\\[-0.8em]
\salt{}            & $0.142 \pm 0.009$ & $0.127 \pm 0.005$ & -- & --\\[0.30em]
\salt{} + \DY & $0.129 \pm 0.008$ & $0.111 \pm 0.005$ &-- & $0.163 \pm 0.029$\\[0.30em]
\salt{} + \DM & $0.132 \pm 0.008$ & $0.116 \pm 0.005$  & $0.119 \pm 0.026$ & --\\[0.30em]
\salt{} + \DY + \DM &$0.126 \pm0.007$ & $0.109 \pm 0.005$  & $0.064 \pm 0.029$& $0.129 \pm 0.032$\\[0.15em]
\hline\\[-0.8em]
\salt{} (on young)     & $0.126 \pm 0.010$ & $0.108 \pm 0.007$ & -- & --\\[0.30em]
\salt{} (on old)       & $0.132 \pm 0.010$ & $0.115 \pm 0.007$ & -- & --\\[0.15em] 
\hline
\end{tabular}
\tablefoot{
$\sigma_{resid}$ is the quadrature sum of 
the additional dispersion needed to obtain a standardization fit with $\chi^2/\mathrm{d.o.f} = 1$
and the 0.055~mag systematic lightcurve fitting error given by \salt{}.
}  
\end{table*}

We tested the stability of the step result by performing four
tests, whose results are summarized in
Table~\ref{tab:summary_systematics}.
For the first test, we rejected SNe~Ia with $c>0.2$. Such red SNe~Ia
are often discarded from cosmological analyses because they are
fainter, leading to biased detection in high-redshift surveys.
Without these, the measured Hubble residual step is 
unchanged, at $\DY = 0.164 \pm 0.029\,\mathrm{mag}$.  
For the second test, we used only SNe~Ia discovered by non-targeted
surveys, i.e., those from SNfactory, LSQ, and PTF, 
as such searches are the most similar to those conducted at
high redshift.  This reduces our sample to $\nSNeUntargeted$ SNe~Ia, and the
resulting brightness offset is $\DY = 0.169 \pm 0.031\,\mathrm{mag}$
-- essentially unchanged.  For the third test, we included
seven SNe~Ia classified as 91T-like since they can be difficult for
higher-redshift surveys to identify.  With these SNe we found
$\DY=0.148\pm0.030\,\mathrm{mag}$.  In this case the
Grubb criterion rejected one 91T-like SN, which is hardly a surprise
given their overluminous nature.  For the fourth test, we checked
the influence of changing the threshold, $\mathrm{LsSFR_{cut}}$, used
to calculate $\PY$ in Eq.~\ref{eq:p(delayed)}. This affects
the fraction of supernovae in our sample classified as younger or older. When
changing $\mathrm{LsSFR_{cut}}$ such that the young fraction ranges
from 60\% (for $\log(\mathrm{LsSFR_{cut}})\sim-11.0$) to 40\% (for
$\log(\mathrm{LsSFR_{cut}})\sim-10.65$), $\DY$ remains
higher than $\sim0.140\,\mathrm{mag}$ and its significance stays
above $\sim5\,\sigma$. 
This test also showed that for our sample, the amplitude and
significance of $\DY$ are maximal when setting $\mathrm{LsSFR_{cut}}$
such that $\sim51\%$ are assigned to the young category. We
reiterate that $\mathrm{LsSFR_{cut}}$ was not tuned for our main
analysis, which followed \citet{rigault_2013} and \citet{rigault_2015}
in splitting the sample exactly in half.
\begin{table*}
\centering
\caption{Summary of how \DY\ depends on perturbations from the main analysis.}
\label{tab:summary_systematics}
\begin{tabular}{l | c c}
\hline\\[-0.8em]
\hline\\[-0.8em]
Choice  & \DY\ [mag] & number of  SNe \\[0.15em]
\hline\\[-0.8em]
remove $c>0.2$           & $0.164 \pm 0.029$& 136 \\[0.30em]
add peculiar SNe         & $0.148 \pm 0.030$& 148 \\[0.30em]
untargeted search only   & $0.169 \pm 0.031$& 114 \\[0.30em]
\hline\\[-0.8em]
$40\%$ with $\mathrm{LsSFR>LsSFR_{cut}}$  & $0.142 \pm 0.030$& 141 \\[0.30em]
$45\%$ with $\mathrm{LsSFR>LsSFR_{cut}}$  & $0.157 \pm 0.029$& 141 \\[0.30em]
$55\%$ with $\mathrm{LsSFR>LsSFR_{cut}}$  & $0.161 \pm 0.029$& 141 \\[0.30em]
$60\%$ with $\mathrm{LsSFR>LsSFR_{cut}}$  & $0.157 \pm 0.029$& 141 \\[0.30em]
\hline
\end{tabular}
\tablefoot{Our baseline analysis is the simultaneous fit of \DY\ with \salt{}
  given in Table~\ref{tab:summary_standardization}. That followed the same
  choices made in \citet{rigault_2013} and \citet{rigault_2015}, for example, splitting the sample 
  in half, removing 91T-like 
  SNe~Ia and not imposing a cut on $c$. 
  The results here explore variants from that baseline.
}
\end{table*}

\subsubsection{Hubble residual dispersions}
\label{sec:magnitude_dispersion}

Another piece of key information for supernova cosmology is the
dispersion around the Hubble diagram. In practice the dispersion
is not explained by measurement uncertainties, and thus represents
missing information that hides unmodeled error. Such errors may
have a systematic component that does not decrease with larger
samples, even for the very large samples expected for future SN~Ia
cosmology surveys.  Reducing the SN magnitude dispersion is thus
one of the best paths for reducing systematic errors, and is of
paramount importance for reaching the accuracy targeted by
future surveys.

The inclusion of $\PY$ as a third standardization parameter
along with $x_1$ and $c$ reduces the weighted RMS (wRMS) of the
standardized SN magnitudes from $0.142\pm0.009\,\mathrm{mag}$ to
$0.129 \pm 0.008\,\mathrm{mag}$.  To test the significance of the
reduction of the dispersion, we remeasured the wRMS while randomly
shuffling the $\PY$ values. We performed 5000 trials, 
and never observed such a low weighted-RMS. 
Hence, with a $p$-value$< 2\times10^{-4}$ we conclude
that using a categorization of SNe~Ia environments using LsSFR
significantly reduces the Hubble residuals dispersion. This remaining
dispersion is still significantly higher than the
$0.077\pm0.011\,\mathrm{mag}$ obtained by the SN twin analysis from
SNfactory \citep{fakhouri_2015}.  This suggests that the SN dispersion
still contains astrophysical effects that are unaccounted for and
that there is still considerable room for improvement. In
Section~\ref{sec:standardization} we examine additional ways to
improve the dispersion using LsSFR.

\subsection{Hubble residual contributions from global mass and LsSFR}
\label{sec:mass_step}

\begin{figure*}
  \centering
  \includegraphics[width=\linewidth]{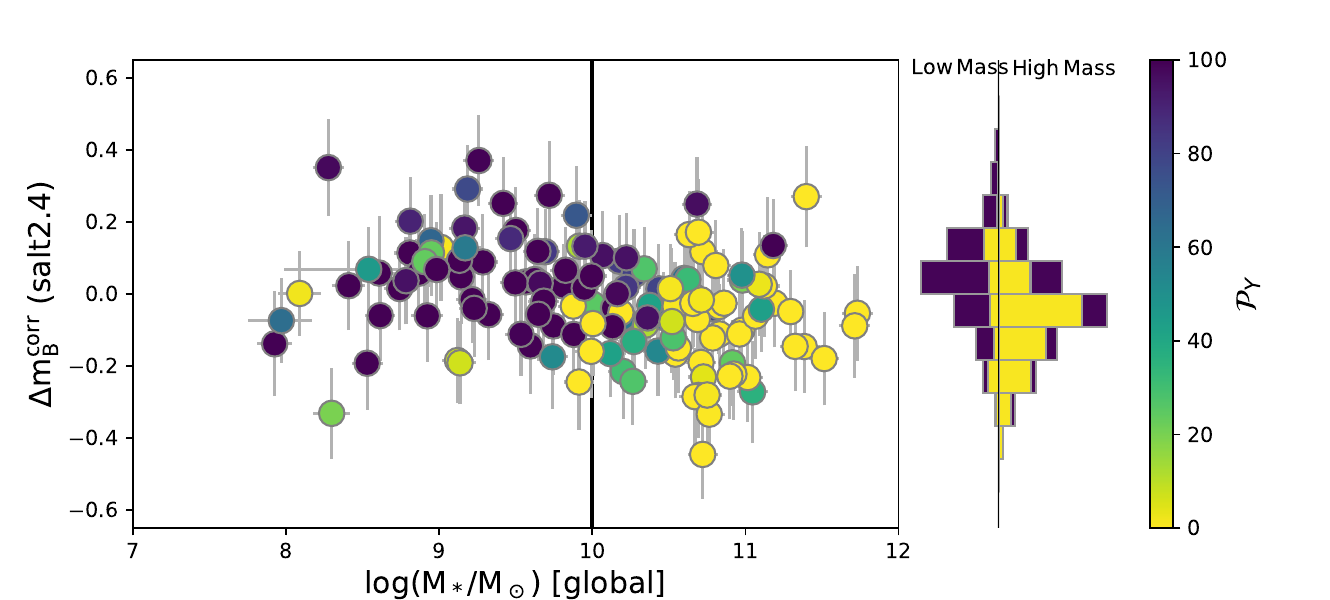}
  \caption{SN~Ia Hubble residuals, $\Delta M_B^{corr}$, calculated from
    a conventional linear standardization using \salt{} 
    lightcurve parameters, as a 
    function of the host-galaxy global stellar mass, $\log(M_*/M_\odot)$. 
    The color code follows that of Fig.~\ref{fig:lssfr_lcparam}. 
    The histograms on the right are $\PM$-weighted
    marginalizations of $\Delta M_B^{corr}$, and thus
    show the contribution of the low  and high global stellar
    mass subsamples.} 
  \label{fig:mass_step}
\end{figure*}

We now explore more deeply the connection between Hubble residual
steps for SNe~Ia segregated by global stellar mass or by LsSFR.
We follow standard practice and classify SNe~Ia as having high host
galaxy stellar mass based on whether their host-galaxy global stellar
mass $\log(M_*/M_\odot)$ is greater than $10\,\mathrm{dex}$. As
with LsSFR and $\PY$, we used a probability distribution,
$\PM$, based on the host stellar mass probability density function.
As in Section~\ref{sec:lssfr_step} for LsSFR, we use this probability
in the computation of the Hubble residual offset between SNe~Ia in
low- and high-mass hosts.  This results in a measured \salt{}-standardized
global stellar mass step of $\DM=0.101\pm0.023\,\mathrm{mag}$,
in agreement with results from the literature
\citep{kelly_2010, sullivan_2010, gupta_2011, childress_2013}. 
This value is significant at $4.3\,\sigma$. The Hubble residuals and
 these fit results are presented in Fig.~\ref{fig:mass_step}.
Alternatively,
mirroring the procedure for LsSFR Section~\ref{sec:lssfr_3rd_param},
we use $\PM$ as a third standardization parameter along
with $x_1$ and $c$. This gives $\DM=0.119\pm0.026\,\mathrm{mag}$,
significant at $4.5\,\sigma$. 

The amplitude and significance of the global stellar mass step is not all that
much smaller than the LsSFR step found in Section~\ref{sec:lssfr_step}.
Some correlation between the two is expected given the known strong
correlation between global sSFR and stellar mass
\citep[e.g.,][]{salim_2014}. We measure a Spearman rank correlation
coefficient of $r_s=-0.63$ between host-galaxy global stellar mass
and LsSFR. This is significant, equivalent to a $9.7\,\sigma$ detection.
This correlation is visible in the histograms of Fig.~\ref{fig:mass_step},
where the younger SNe~Ia favor lower-mass hosts while the older
SNe~Ia favor hosts of higher mass. That still leaves about 25\% of the SNe~Ia 
that are classified as young in a high-mass host or old in a low-mass host. 
This suggests that the global stellar mass step
might, at least partially, be a consequence of the LsSFR step.

To test this hypothesis, we simultaneously fit for the global stellar mass step,
$\DM$, and the LsSFR step, $\DY$, along with the
standardization coefficients for the \salt{} lightcurve parameters.
We find $\DY=0.129 \pm 0.032\,\mathrm{mag}$, a $4.0\,\sigma$
detection for the LsSFR step, versus $\DM= 0.064 \pm
0.029\,\mathrm{mag}$, a $2.2\,\sigma$ detection for the mass
step. The resulting $\mathrm{wRMS}=0.127 \pm 0.008\,\mathrm{mag}$ 
is similar to what is obtained when fitting only the LsSFR step
(see Table~\ref{tab:summary_standardization}).

We draw three conclusions from these results:
(1) Because the amplitude and significance of $\DY$ are
greater than those of $\DM$, the driving environmental
dependency seems to be the SN~Ia age. This statistical
result supports the physical argument that the
LsSFR, as a tracer of the fraction of young stars at the SN
location, is more closely connected to the SN progenitor than
is the total stellar mass of the host galaxy. Put another way, approximately
70\% of the variance from the stellar mass step is due to an underlying
dependence on progenitor age as inferred from the local environment.
(2) Because $\DY$ remains quite significant when including
$\DM$, SN~Ia standardization using the global stellar mass step leaves
residual systematic errors.
(3) Because the amplitude of $\DM$ remains non-negligible 
($2.2\,\sigma$), progenitor age may not reflect the full
SN~Ia environmental dependency.

\begin{figure}
  \centering
  \includegraphics[width=\linewidth]{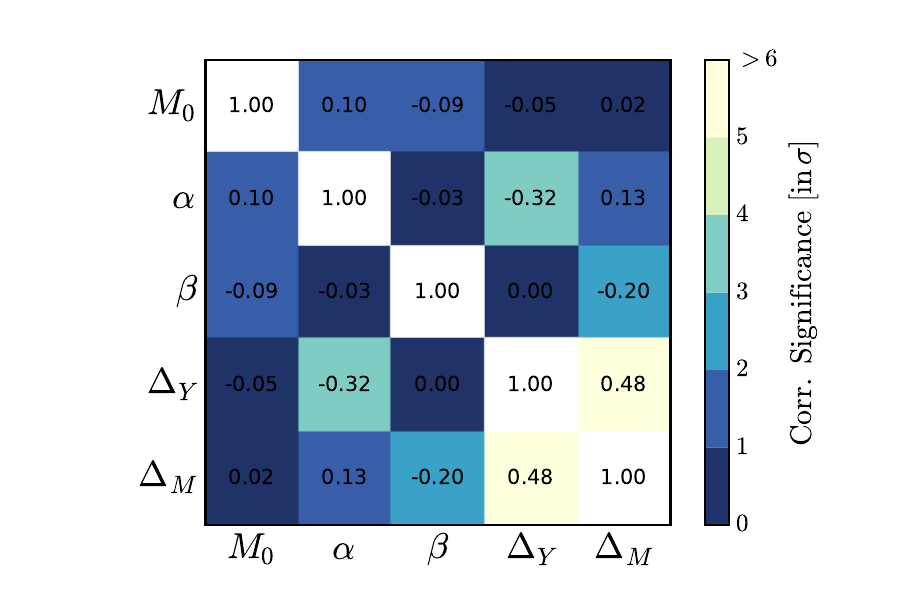}
  \caption{Correlation matrix between the coefficients of the
    4-parameter standardization (\salt{} $x_1$ and $c$, along with
    $\PY$ and $\PM$). The values within the matrix are
    the correlation coefficients. The color of the matrix elements
    represent the significance of this correlation coefficient.} 
  \label{fig:correlation_matrix}
\end{figure}

The correlation matrix between the absolute magnitude, $M_0$, the
SN lightcurve stretch standardization coefficient, $\alpha$, the
SN lightcurve color standardization coefficient, $\beta$, $\DY$
and $\Delta_M$ for the original simultaneous standardization is
shown in Fig.~\ref{fig:correlation_matrix}.  \footnote{For calculation
of the standardization, $\PY$ and $\PM$ are translated
to be centered around 0 -- ranging from $-0.5$ to $+0.5$ -- such that
the correlation with $M_0$ is consistent with 0 by construction.
The same is true for $x_1$ and $c$. This transformation has no
effect on the derivation of  $\DY$ and $\DM$.} This
matrix summarizes some of our results. The correlation of $r=-0.31$
between $\DY$ and $\alpha$ represents the correlation between
$x_1$ and $\mathrm{LsSFR}$ discussed in
Section~\ref{sec:result_lssfr_stretch}. The correlation between
$\DY$ and $\DM$ reflects the correlation between the
host-galaxy global stellar mass and the local sSFR discussed above.  The
lack of correlation between $\DY$ and $\beta$ reflects our
finding in Section~\ref{sec:result_lssfr_stretch} that the progenitor
age does not significantly influence the lightcurve color measured
by \salt{}.  Finally, the correlation of $r=-0.20$ between $\DM$
and $\beta$ might be a sign that the global stellar mass step carries additional
information, for example, about progenitor metallicity \citep{childress_2013}
or amounts or properties of dust.

\subsection{Standardization by subpopulation}
\label{sec:standardization}

The difference between the standardized magnitudes of younger and
older SNe Ia calls into question the uniformity of the stretch and
color standardization process.  To test this, we independently
standardize SNe~Ia from each group to compare their standardization
coefficients, $\alpha$ and $\beta$. (The $M_0$ for each subpopulation
absorb the LsSFR step).  For these fits, as for the KS test,
we categorize the $\nSNeYoung$ SNe~Ia with $\log(\mathrm{LsSFR}) > -10.8$ 
as young and the
$\nSNeOld$ SNe~Ia with $\log(\mathrm{LsSFR}) \le -10.8$  
as old.  Changing this partitioning
does not significantly affect our results.  The differences in the
standardization coefficients are presented in
Table~\ref{tab:delta_param_delayed}.

\begin{table*}
\centering
\caption{Variation of standardization coefficients between young and old subpopulations.
}
\label{tab:delta_param_delayed}
\begin{tabular}{l | c c}
\hline\hline\\[-0.8em]
Standaridzation & Change between   & Change between \\
Coefficient     & young and old    & young and old for $c<0.2$\\[0.15em]
\hline\\[-0.8em]
$\Delta M_0$ & $+0.132 \pm 0.024$  ($5.5\sigma$)  & $+0.140 \pm 0.025$  ($5.6\sigma$)\\[0.30em]
$\Delta \alpha$ & $+0.013 \pm 0.024$  ($0.6\sigma$)  & $+0.006 \pm 0.025$  ($0.3\sigma$)\\[0.30em]
$\Delta \beta$ & $+0.261 \pm 0.246$  ($1.1\sigma$)  & $-0.158 \pm 0.389$  ($0.4\sigma$)\\[0.30em]
$\Delta \sigma_{int}$ & $-0.009 \pm 0.010$  ($0.9\sigma$)  & $-0.010 \pm 0.011$  ($0.9\sigma$)\\[0.15em]
\hline
\end{tabular}
\end{table*}

We find a number of interesting results when standardizing the
subpopulations independently. 

The first is that the $\alpha$ standardization coefficient, which
accounts for the Phillip's ``brighter-slower'' relation, is consistent
between the two age groups. This is despite our findings in
Section~\ref{sec:result_lightcurve} that the two populations span
different ranges in $x_1$ and that, overall, LsSFR and $x_1$ are
strongly correlated. Similarly, we find no difference in
the color correction coefficient, $\beta$. This is in contrast with
the  significant differences in $\beta$ when dividing by global
host galaxy properties \cite{sullivan_2010}, although that difference
was found to depend strongly on the few reddest SNe~Ia in their sample.

As expected, the LsSFR step translates into a difference in the
mean absolute magnitudes.  The difference is $\Delta M_0=0.132\pm0.022$~mag,
significant at $5.5\,\sigma$, and remains when removing the reddest
($c>0.2$) SNe. The fact that the elimination of red SNe~Ia has
such little effect suggests that the LsSFR bias is not driven by
differences in SN~Ia intrinsic colors, dust extinction, or the
tension between the two that is inherent when a single parameter
is used to correct for both effects.

When allowing for independent standardizations (using $x_1$ and $c$)
for each population, the
younger population exhibits the smallest weighted RMS yet seen in
this analysis: $\mathrm{wRMS}=0.126 \pm 0.010\,\mathrm{mag}$.
This compares with $\mathrm{wRMS}=0.142\pm0.009\,\mathrm{mag}$
before accounting for any environmental biases, and $\mathrm{wRMS}=0.129
\pm 0.008\,\mathrm{mag}$ when fitting the full sample for the LsSFR
step. This wRMS based on the \nSNeYoung\ younger SNe~Ia is still
$\sim3\,\sigma$ higher than the twin SN dispersion of $0.077 \pm 0.011\,\mathrm{mag}$
determined for 55 SNe~Ia in \citep{fakhouri_2015}, and $\sim2\,\sigma$
higher than the dispersion of $0.075 \pm 0.018\,\mathrm{mag}$
determined by \citep{kelly_2015} using 11 SNe~Ia having locally
UV-bright environments.

The older population has $\mathrm{wRMS}=0.132 \pm 0.011\,\mathrm{mag}$,
and therefore has a slightly worse standardization (by $\sim1\,\sigma$)
than the younger population (see additional tests in
Section~\ref{sec:impact_of_onesided}).   This supports the claims of
\cite{rigault_2013,rigault_2015}, \cite{childress_2014} and
\cite{kelly_2015} that SNe~Ia from younger progenitors are more
favorable for cosmological analysis since for them the hidden
astrophysical systematics that remain are smaller. Advantages of
this kind will be critical in the era of new large surveys, where
statistics will no longer be an important limitation.

\section{Cross-checks and comparisons}
\label{sec:discussion}

Here we examine other factors that could conceivably influence
our results.  It is first important to establish the context and
set the scale.  We are studying an effect, not a parameter
tied to a fundamental physical model, so the only thing that matters 
for this section is whether or not the appearance of this effect could itself
be induced by systematic errors.  Our uncertainty is
$\sim0.03$~mag, and our signal is $\sim0.16$~mag, therefore,
systematic uncertainties of order $\sim0.05$~mag would be needed
to substantially change our view of the LsSFR effect (i.e., potentially
moving the measured offset by more than $2\,\sigma$, or, equivalently,
potentially decreasing the significance of the measured offset below
$3\,\sigma$).  A systematic error of this magnitude is more than $5\times$
less stringent than the level of systematic error control required
for the measurement of cosmological parameters.  Moreover, most
sources of systematic error that must be accounted for in cosmological
measurements cancel out in the analysis here.  For example, there
are negligible K-correction errors (solely for the
  host mass measurements, since the rest of the analysis is
  spectroscopic) 
or evolution effects since our redshift
range is so small.  Calibration zero-point or color errors cancel
out because an overall calibration is performed in the same way
for SNe~Ia in different types of host galaxies. We now consider
additional possible effects.

\subsection{Signal Dilution}
\label{sec:impact_of_dilution}

LsSFR is not a property intrinsic to SNe~Ia, but rather a
means of attempting to sort them by some intrinsic property, which
we think is related to progenitor age. Therefore, the true LsSFR
step in Hubble residuals represents a lower limit for a given
standardization method since any error in sorting by a property
such as age decrease the measured size of the step. Line-of-sight
projection, use of poor methods or data quality for measuring local
star formation or masses, etc., move some SNe~Ia to the wrong
side of LsSFR$_{cut}$, thereby reducing $\DY$.  Quantitatively, if
the mis-categorization fraction is $\xi$ then the size of the step
that is measured decreases by $2\,\xi$.  (Also, by this argument,
a better metric than LsSFR would increase $\DY$.) Given that we
already have a significant measurement of the LsSFR bias, errors of this
nature in the current measurements cannot eliminate the LsSFR bias
we have observed.

\subsection{Robustness of the LsSFR bias to host galaxy subtraction}
\label{sec:impact_of_hostsub}

In this section we examine whether errors in host-galaxy
subtraction might change our measurement of the SN brightnesses in
a way that could mimic the LsSFR bias.  We first note that in the
presentation of our host-galaxy subtraction algorithm in
\citep{bongard_2011}, two nominally challenging cases, one having
a galaxy nucleus and strong spiral arms, the other an edge-on spiral,
were presented and the resulting residuals demonstrated to be 
clean. Visual inspection of the modeling residuals also shows no
host-galaxy subtraction issues in the sample studied here. Therefore, we
have no a priori reason for suspecting issues with host-galaxy
subtraction.

Since host stellar mass appears in the denominator of LsSFR, and
stellar mass is derived from the host galaxy light, for host
subtraction errors to generate a false LsSFR bias requires preferential
oversubtraction in the cases where the host light is fainter than
average (giving higher LsSFR), or undersubtraction in the cases
where the host light is brigher than average (giving lower LsSFR).
In the process, the dispersion for the SNe~Ia would be required to
substantially improve --- to 0.103~mag and 0.115~mag for
higher and lower LsSFR, respectively --- over the canonical
$\sim0.15$~mag generally found for standardization using SALT. It
is difficult to imagine a scenario in which such an anti-correlation
of host subtraction errors and a simulataneous substantial improvement
in the Hubble residuals could be produced.

Nonetheless, we can examine the question of whether host
subtraction errors could be large enough to matter here.  
To do this we measured the host-galaxy brightnesses at the SN
locations and then remeasured the LsSFR step after eliminating from the
sample SNe with various levels of high host-galaxy background. We
find that changes in the size of the LsSFR step are small and well within the
uncertainties. Even for an extreme case, in which we require that
the host-galaxy background to be no more than 2\% of the SN maximum brightness
--- a requirement that eliminates half of our sample, we still find
a LsSFR step of $-0.131\pm0.027$~mag, which is consistent with that
from our full sample.

The flux from H$\alpha$ is too small (less than a few percent
even for our strongest line) to affect the broadband photometry
used when fitting SNe~Ia lightcurves, so the possibility of Hubble
residual errors due to mis-subtraction of H$\alpha$ need not be
given any further consideration. The measurement of H$\alpha$ itself,
used in the numerator of the LsSFR measurements, is determined
relative to the surrounding galaxy continuum, so is immune to offsets
in the baseline flux.

From these considerations we conclude that, for our data
processing and analysis, host-galaxy subtraction errors --- either
in the numerator or demoninator --- are too small to impact the
measurement of the LsSFR bias.

\subsection{Robustness of the LsSFR bias to dust}
\label{sec:impact_of_dust}

There are two ways host dust extinction errors could enter into
our measurement of the LsSFR bias: first through the standardization
of the SN brightnesses based on their observed colors, and second,
through the measurement of LsSFR.

There is evidence that variations in dust properties impact
the standardization of SNe~Ia \citep[][and references therein]{huang_2017}.
But for the analysis here, any such systematic variations can be
considered to be part of the signal of how of host-galaxy environments
impact the standardization of SNe~Ia. That is, if SNe~Ia have
different dust properties due to differences in their local
environments, that too is likely related to age since dust formation
and subsequent reprocessing is directly tied to star formation.
Therefore, while systematic errors in the extinction correction of
cosmological SNe~Ia is important, for our work it is not a source
of systematic uncertainty.

Next, since our H$\alpha$-based LsSFR and our global and host-galaxy
local photometry is not corrected for dust extinction, we revisit
the extent to which this might affect our results. While both the
numerator (the local SFR) and denominator (the local
 mass) in our LsSFR indicator are suppressed 
by dust, incomplete cancellation is expected for two reasons.  First, in
galaxies the dust extinction curve is flatter for stars than for
HII regions \citep[e.g.,][]{calzetti_2000,kreckel_2013}, so
the local SFR as measured from H$\alpha$ is suppressed relative 
to stellar mass as measured from star light. In addition, dust-reddened $g-i$ results in a
higher estimated mass-to-light ratio when calculating stellar
masses, offsetting some of the effect of extinction.

To better quantify this effect, we simulated expected amounts of
dust based on the SFR versus $E(B-V)$ and stellar mass versus
$E(B-V)$ relations given by \cite{battisti_2016}, including the
scatter about the mean relations. These are consistent
with global galaxy SFR and dust trends as well \citep{brinchmann_2004}.
These trends show that dust increases along the locus where both
stellar mass and star formation are increasing. The net effect is
to compress and slightly distort the measured LsSFR relative to the
true LsSFR.  Our LsSFR step analysis uses a threshold,
$\mathrm{LsSFR_{cut}}$, selected to divide our sample in half, so in
the mean these effects are not expected to impact our categorization
of SNe~Ia between younger and older progenitors.

It is therefore not surprising to find that, after statistically
correcting our LsSFR measurement for dust attenuation based on the
\cite{battisti_2016} relations, the LsSFR step, \DY, drops by only
a fraction of the given error ($\sim0.015\,\mathrm{mag}$). Thus,
while our extincted LsSFR may be slightly distorted, modeling of
the effect indicates this has negligible impact on our main results.

\subsection{Independence of LsSFR from metallicity}
\label{sec:impact_of_metallicity}

As a further check on metallicity dependence, we find
that LsSFR values in our sample are somewhat correlated with
host-galaxy global gas-phase metallicities for the \nOH\ galaxies
having both LsSFR from this study and gas-phase metallicities from
\cite{childress_2013b}. The Spearman correlation coefficient is
$r_s = -0.25$, which has a significance of
$2.0\,\sigma$. This 
subsample is primarily restricted to the younger SN population since
ionizing stars are needed to produce the emission lines used to
measure gas-phase abundances. Thus, we can't fully answer the
question of the potential impact of metallicity on the sample as a
whole. But since it is for star-forming galaxies like these that
\cite{laralopez_2013} found some metallicity trend, it is likely
that this trend is more of an upper limit to the effect of
metallicity on LsSFR for our overall sample.

\subsection{Local stellar mass bias}
\label{sec:impact_of_local_mass}

It is also interesting to look at whether there is a Hubble residual
step when categorizing SNe~Ia by the local stellar mass. We find
that when splitting the current sample at the median local stellar
mass value of $\log(\mathrm{M_*/M_{\odot}})=8$, the
SNe~Ia with low local stellar mass are
$0.059\pm0.024\,\mathrm{mag}$ fainter than those with high local
stellar mass. After accounting for the LsSFR step this falls to 
$0.021\pm0.022\,\mathrm{mag}$ suggesting that the local mass
step simply is due to the correlation between local mass and LsSFR
($r_s=-0.24;\ 2.9\,\sigma$).

As expected from the structure of galaxies, in our sample
there is no correlation between local and global stellar mass
beyond that from the limit that local stellar masses can not
exceed global stellar masses.

\subsection{Robustness when splitting the sample by stretch or color}
\label{sec:impact_of_stretch_color_cut}

Another way to test for potential non-uniformity in the
standardization follows \cite{sullivan_2010}, who examined the
variation of the brightness offset between SNe~Ia in low- and
high-mass hosts when splitting the sample at $c=0$ or $x_1=0$.  To
perform these tests, we measure the corresponding values of $\DY$
after standardization, as in \cite{sullivan_2010}.
Therefore the results of these tests are to be compared with the
$\DY=0.125 \pm 0.023\,\mathrm{mag}$ presented in
Section~\ref{sec:lssfr_step}:
\begin{itemize}
\item For the 59 having $c>0$ $\DY=0.134\pm
  0.041\,\mathrm{mag}$, compared to $\DY=0.112\pm  0.027\,\mathrm{mag}$ for the
  remaining 82 having $c<0$;
\item For the 82 having $x_1>0$ $\DY=0.151\pm 0.028\,\mathrm{mag}$ compared to
  $\DY=0.110\pm  0.040\,\mathrm{mag}$ for the remaining 59 having $x_1<0$
\end{itemize}

It is apparent that on each side of these dividing lines
the SNe~Ia show a significant LsSFR bias. Moreover, the size of the
LsSFR bias is consistent between these subsets.
This result strengthens our conclusion from
Section~\ref{sec:standardization} that the brightness offset between
the younger and older SNe~Ia cannot be fixed by
simply modifying the linear standardization based \salt{} parameters.

\subsection{Robustness when fitting non-linear stretch and color relations}
\label{sec:impact_of_broken}

\cite{rubin_unity_2015, scolnic_2016} presented evidence that standardization
using $x_1$ and $c$ is improved by using non-linear relations.
This motivates an examination of the potential impact
of non-linear standardization on
the LsSFR bias.  We applied the UNITY framework of \cite{rubin_unity_2015}
and found that $\DY$ is just as strong when broken-linear standardization
is allowed. Also, we find almost no covariance between the broken
standardization coefficients and $\DY$, consistent with the results
given in Section~\ref{sec:results}.

\subsection{Physically motivated outlier rejection}
\label{sec:impact_of_onesided}

In our main analysis we applied the Grubb criterion to identify
potential outliers. This rejection, which was blind and based on a
two-sided test, found no outliers.  However, while our local technique
is an improvement over global techniques in isolating the stellar
environment of each SN, incorrect categorization is possible due
to projection along the line of sight. If a younger SN were projected
onto a region with low LsSFR it would be misclassified as a older SN
that is too faint.  Conversely, if a older SN were projected onto a
region with high LsSFR it would be misclassified as a younger SN that
is too bright.  As discussed in detail in \cite{rigault_2013}, since
older stars develop higher velocities and have more time, they are
more likely to move away from their original environment. This
motivated a test for evidence of missclassifications.

We revisited the LsSFR step and per-population standardization using
a one-sided Grubb criterion, thereby allowing rejection of unexpectedly
bright and young or faint and old SNe.  Doing so finds only one
case: SNF20060912-000, which is categorized as young but found to
be too bright when we perform \salt{} standardization of the younger
population (see Section~\ref{sec:standardization}). After this SN~Ia is
rejected, the dispersion for
the young population falls to $\mathrm{wRMS} = 0.120\pm 0.011\,\mathrm{mag}$,
which is $2\,\sigma$ smaller than for the standardization using
only the old subpopulation. Of course such changes are guaranteed
when applying one-sided rejection; that they are small and only one
SN was affected suggests that projection effects are not an important problem
for the LsSFR indicator.

\subsection{Alternative test of the reduction of the global stellar mass step}
\label{sec:mass-step_randomization}

To verify that the reduction of the global stellar mass step
presented in Section~\ref{sec:mass_step}
is caused by the inclusion of information about the progenitor
age, and not any fourth parameter, we reran the simultaneous fit using
$x_1$, $c$, $\PY$ and $\PM$, each time randomly shuffling the $\PY$
values.  For these 5000 randomizations, the recovered global stellar mass step
peaks at $\Delta_M=0.119\,\mathrm{mag}$ and has a standard deviation
of $0.002\,\mathrm{mag}$. Randomly finding a reduction
in the global stellar mass step fluctuating as low as $0.064\,\mathrm{mag}$ is
thus excluded at $\gg5\,\sigma$. We consequently conclude that the
global stellar mass step is at least partially caused by the LsSFR
offset.

\subsection{Comparison to \cite{rigault_2013}}

This work extends our first analysis of the environments surrounding
individual SNe~Ia, where we used the local SFR, LSFR, to probe
progenitor properties and notably its age \citep{rigault_2013}.
However, as discussed above, the LsSFR provides important additional
information by effectively normalizing by the SN rate contribution
from older progenitors at the SN location.  The LsSFR and the LSFR
indicators are positively correlated at a significance of $12\,\sigma$
in our data set. For SNe in common with \cite{rigault_2013}, about
25\% change their environmental classification when using LsSFR
rather than LSFR.  Classification shifts from the \cite{rigault_2013}
$\mathrm{Ia\epsilon}$ category to large $\PY$ arise from moderate/low
SFR cases within regions with low local stellar mass.  There, even
a small amount of star formation is enough to strongly favor a young
progenitor given the lack of an underlying old stellar population.
Such cases typically have a $\PY\sim50$--70\%, reflecting the larger
errors when both SFR and local stellar masses are low.  Classification
shifts from the \cite{rigault_2013} $\mathrm{Ia\alpha}$ category
to low $\PY$ correspond to moderate/high SFR values (slightly above
the \citealt{rigault_2013,rigault_2015} cut of $-2.9\,\mathrm{dex}$)
that are superimposed on regions with large local stellar mass.
Many of these may correspond to the false-positive category that was
discussed in \citealt{rigault_2013} and \citealt{rigault_2015},
because the chance of having an older progenitor misassociated with
star formation is increased.

It is interesting to make a quantitative comparison between our new
results using LsSFR with our previous results from \cite{rigault_2013} using the local star formation rate, LSFR. The
\cite{rigault_2013} sample had roughly half the size of the current
sample.  Using \saltto{}, as in \citet[][see also erratum]{rigault_2013}, the LSFR step, measured after
  standardisation as in section~\ref{sec:lssfr_step}, is $0.063 \pm 0.029\
\mathrm{mag}$, whereas the LsSFR step is $0.110 \pm
0.033\,\mathrm{mag}$.
When using \salt{} instead, the LSFR step decreases to $0.045 \pm
0.029\ \mathrm{mag}$ while the LsSFR step is
$0.088\pm0.030\,\mathrm{mag}$. This 
is when rejecting the three bright SNe~Ia that appeared to be
misclassified by LSFR in \cite{rigault_2013}, however,
using LsSFR these cases appear to be correctly classified and when
they are included we find an LsSFR step of
$0.110\pm0.030\,\mathrm{mag}$ when using \salt{}.  This
lends further support for LsSFR being a better age discriminator.
As discussed in Section~\ref{sec:impact_of_dilution}, since a
mis-categorization fraction of $\xi$ leads to a measured step
decreased by $2\,\xi$, the difference between the LSFR and the LsSFR
steps are consistent given the aforementioned $\sim25$\% of SNe~Ia
classified differently and assuming that the LsSFR classification
is more correct.

The biggest changes between \pkg{SALT2.1} and \salt{} are
in the mean SN~Ia spectral model and in the color correction function
\citep[see Figs.~2 and 3 of][]{betoule_2014}. The R13 SNe~Ia proximate
to active star formation were redder by $0.036\pm0.017$, so it is
not surprising that changes in the \pkg{SALT} color model would have
an effect.  However, our finding that LsSFR is strong even when
using the new \salt{}, suggests that simply retraining 2-parameter
lightcurve models does not remove environmental dependencies. In
\cite{kim_2013} we found that a 4-parameter lightcurve model could
reduce the bias with stellar mass, and in \cite{nordin_2018} we
found that UV data can also reduce the bias with LsSFR. These results
indicate that improved lightcurve fitting and standardization methods
should be pursued in concert with studies of SN~Ia environments.

\section{Progenitor age bias and SN cosmological measurements}
\label{sec:cosmo_bias}

We have found that supernovae associated with younger progenitors
are significantly fainter than those associated with older progenitors
after $x_1$ and $c$ standardization. Therefore, if the fraction of
younger SNe changes between SN samples or as a function of redshift,
the average SN~Ia magnitude will not be standard and the derived
cosmology will be biased. This may already be an issue for the
measurement of the Hubble constant, $H_0$, due to differential
selection between Hubble flow and calibrator SNe~Ia environments
\citep{rigault_2015}.  The impact on $H_0$ has not been fully
resolved \citep[][]{rigault_2015,jones_2015, riess_2016}, and so
we plan to explore this further in a separate study using our new
LsSFR environmental indicator.  A redshift dependency could be even
more problematic, as it would affect the estimation of the dark
energy equation of state parameters, especially measurement of
its time variation, $w_a$ ($w=w_0 + w_a z/(1+z)$).  We here estimate
the expected redshift bias, and discuss ways to account for it.

\subsection{Redshift evolution of progenitor age}
\label{sec:redshift_evolution}

As outlined in \cite{rigault_2013}, the primary reason to be concerned
about an age bias is that the mean fraction of young stars is known
to strongly evolve with redshift.  The sSFR is an order of magnitude
greater at $z=1.5$ than at $z=0$ \citep[see][for a
review]{madau_2014}. The theoretical expectation is that
$\mathrm{sSFR}\propto(1+z)^{2.25}$ \citep{dekel_2009}, while
observations give an even steeper dependence of $(1+z)^{2.8\pm0.2}$
\citep{tasca_2015}.

In decompositions of the SN~Ia progenitor delay time distribution
into younger and older categories, it is assumed that statistically the rate of younger/prompt progenitors
is proportional to the SFR while the rate of older/delayed progenitors is
proportional to the host stellar mass, $M_*$
\citep{mannucci_2005, scannapieco_2006}.
In such a schematic
model, the ratio between younger and older progenitors would be
proportional to the sSFR. LsSFR would reflect this
ratio in the vicinity of each SN~Ia.

Consequently, denoting the evolving fraction of young and old SNe~Ia
as $\delta(z)$ and $\psi(z)$, respectively, as in \cite{rigault_2013},
gives the redshift evolution of their ratio as:
\begin{equation}
\label{sec:ssfr_evolution}
  \frac{\delta(z)}{\psi(z)} \equiv \mathrm{LsSFR}(z) = K \times (1+z)^{\phi},
\end{equation}
and consequently,
\begin{align}
\label{sec:p(delayed)_redshift}
 \delta(z) &= \left(K^{-1}\times(1+z)^{-\phi} +1\right)^{-1}, \mathrm{or}
  \nonumber\\
  \psi(z) &= \left(K\times(1+z)^{+\phi} +1\right)^{-1},
\end{align}
which require a 50--50 split between young (prompt) and old (delayed) progenitors sets the
coefficient $K=0.87$ for our $z\sim0.05$ sample when using the value
$\phi=2.8$ found by \cite{tasca_2015}. For the redshift range
spanned by most of the existing SN~Ia cosmology samples, this
relation is similar to the approximation sSFR $\propto 10^{0.95\times
z}$ made in \cite{rigault_2013}.  Then, if we assume that the brightness
offset between younger and older populations, $\DY$, is a
constant with redshift --- as expected if this effect 
arises from the physics of the progenitors --- the mean standardized
magnitude of SNe~Ia at maximum light can then be written as:
\begin{align}
  \label{eq:HR_z}
  \langle M_B^{corr}\rangle (z) &=  \delta(z) \times \langle
                                  M_B^{corr} \rangle_{prompt} +
                                  \psi(z) \times \langle M_B^{corr} \rangle_{delayed} \nonumber\\
                                &= \langle M_B^{corr} \rangle_{prompt}
                                  - \psi(z) \times \DY.
\end{align}
Thus,  as $\psi(z)$ tends toward zero with increasing redshift, the
average SN~Ia magnitude, $\langle M_B^{corr}\rangle$, tends toward
the average magnitude of the young/prompt population $\langle M_B^{corr}
\rangle_{prompt}$.

\begin{figure}
  \centering \includegraphics[width=\linewidth]{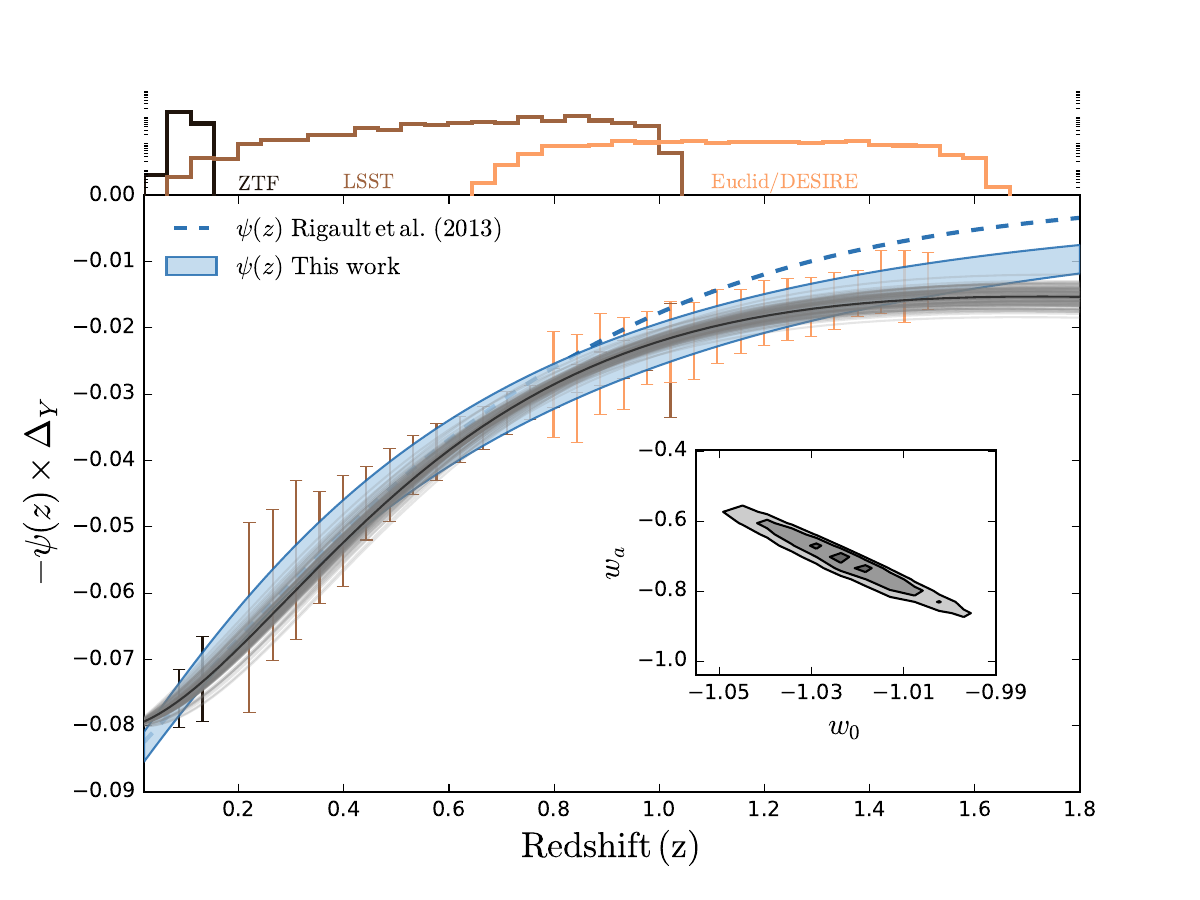}
  \caption
{Evolution of the LsSFR bias, $-\psi(z)\times \DY$, with
redshift, as predicted by Eq~\ref{eq:HR_z}. The blue band shows
the expected redshift evolution of $-\psi(z) \times \DY$
when accounting for uncertainties on $\phi$. 
The result of using the $\psi(z)$
parametrization in \cite{rigault_2013} is shown as a blue dash line
for comparison (see also \citealt{childress_2014}).  To quantify
the impact of the LsSFR bias, we simulate data following the expected
SN rate and data quality of ZTF, LSST and \Instru{Euclid}/DESIRE (see text).
The expected numbers of SNe~Ia are shown above the main plot on a
logarithmic scale, color coded by sample. The expected binned
uncertainties on the measurements of $-\psi(z)\times \DY$
are shown in the main plot, following the same color code.  Randomly
drawn MCMC realizations of the fitted magnitude difference between
a flat $w_0\,w_a$CMD and a flat $\Lambda$CDM are shown as a grey band.
The black line is the best fit value. The extracted $w_0,w_a$
posterior distribution is displayed in the inset plot.  This figure
illustrates that the LsSFR bias would be mistakenly attributed to
the key properties of dark energy that future surveys hope to
measure.}
  \label{fig:environmental_bias}
\end{figure}

We now use Eq.~\ref{eq:HR_z} to estimate the resulting bias on
the measurement of the dark energy equation of state parameters,
$w_0$ and $w_a$.  Fig.~\ref{fig:environmental_bias} illustrates
the results.

For this example, we assume the expected SN~Ia numbers
from the LSST deep-fields, \Instru{Euclid}/DESIRE  \citep[as given by][]{astier_2014}
and ZTF SN sample (private communication).   
The exact SN numbers are irrelevant for the model,
but using them provides an idea of the expected ability for these
surveys to be affected by the redshift evolution of the LsSFR bias.
In each $\Delta z=0.2$ redshift bin we simulate the 
$-\psi(z) \times \DY$ term of Eq.~\ref{eq:HR_z}, which is
the portion that varies with redshift.
This particular model also assumes:  
\begin{enumerate}
\item 
that $\DY=0.16\,\mathrm{mag}$, as in Section~\ref{sec:standardization},
and is independent of redshift. The limit of this assumption
is that the LsSFR bias could depend on the
mean age of the older population.  Then $\DY$ would be expected
to decrease as a function of redshift since higher-$z$ stars are
younger. However, this would amplify the cosmological biases.

\item 
that the ratio of younger to older SNe~Ia follows
Eq.~\ref{sec:ssfr_evolution}.  This implicitly assumes no survey
selection efficiency against either LsSFR category. In 
practice, selection effects are inevitable, as discussed
in Section~\ref{sec:lssfr_redshift}, and will need to
be taken into account for real data.

\item 
that the data-quality all along the redshift range covered by these
surveys is similar to that for the SNe~Ia in this paper. That is,
we assume that the error on $\DY$, based on our measurement
error on $\DY$ of $\sim0.03\,\mathrm{mag}$ for \nSNe\ SNe,
goes as $0.03\times \sqrt{\mathrm{\nSNe/N_{SNe}}}$, where $\mathrm{N_{SNe}}$
is the number of SNe~Ia for a given survey in a given redshift bin.
Such data quality is expected for next-generation surveys
\citep{kim_2015}.

\item 
no mass-step correction. The redshift evolution of the relation
between the host-galaxy global stellar mass and the sSFR is complex
\citep[e.g][]{faber_2007,bauer_2013, johnston_2015}. Furthermore,
global stellar masses evolve across any fixed mass threshold
as galaxies grow and merge.  Therefore, understanding how to correct
for any global stellar mass step remaining after correction for the LsSFR bias is
far from trivial, and is beyond the scope of our simple model.

\end{enumerate}

We fit the simulated $\psi(z) \times \DY$ with the calculated
brightness differences between a nominal flat $\Lambda$CDM model
and a flat, $w_0\,w_a$CDM model. We fixed $\Omega_M$ to the \cite{planck_cosmo_2015}
value, and for the flat $\Lambda$CDM $w_0\equiv -1$ and $w_a\equiv
0$. Therefore, for the flat w0waCDM model $w_0$ and $w_a$ are the
only free parameters.  The resulting contours for the recovered
$w_0, w_a$ values from MCMC fitting are shown in
Fig.~\ref{fig:environmental_bias}. We find very significant shifts
of $\Delta w_0=0.03\pm0.01$ and $\Delta w_a=0.7\pm0.1$. These
shifts completely ruin the measurement of dark energy properties.
Without a $z>1$ sample, the bias on $w_0$ increases to $\Delta w_0
\sim 0.05$, as found in \cite{rigault_2013}.

We also studied the size of the effect after mimicking the
current practice of standardizing SNe~Ia including a redshift-independent
step based on host-galaxy global stellar mass. For our dataset this
form of standardization still leaves a LsSFR step of $0.076 \pm
0.022$ mag. If we then allow this step to evolve, as above, serious
biases of $\Delta w_0=0.03\pm0.01$ and $\Delta w_a=0.3\pm0.1$
remain.

These examples illustrate the paramount importance of accurately
accounting for astrophysical biases that may evolve with redshift
or develop due to survey selection effects. The brightness offset
between the younger and older populations, combined with the expected
evolution of their ratio based on the well-known sSFR-redshift trend
badly biases the determination of the dark energy equation of
state parameters if not taken into account. And, as the residual
dispersion illustrates, more systematic biases could be lurking in
the SN~Ia data. There are some indications, that improved lightcurve
fitters could help, such as the reduced LSFR bias in going from
\pkg{SALT2.1} to \salt{}, or the multi-component lightcurve model
of \cite{kim_2013} and with spectrophotometry, it is possible
that twinning addresses these biases.  Encouragingly, as described in the next
section, measuring LsSFR at high redshift appears feasible.

\subsection{Measuring LsSFR at high redshift}
\label{sec:lssfr_redshift}

For the present analysis we have used H$\alpha$ to measure the local
SFR. Such data may be rare for high-redshift SNe~Ia, but could be
obtained with the integral field spectrographs planned for \Instru{JWST} and
\Instru{WFIRST}. Deep imaging is more commonly available because it is built
up naturally over the course of long-duration wide-field SN searches.
As an alternative to H$\alpha$, the specific star formation rate
can be derived from photometric SED fitting \citep[e.g.][]{conroy_2009}.
In SED modeling, the sSFR is usually referred to as the Scalo $b$
parameter -- the current-to-past star formation ratio
\citep[e.g.][]{scalo_1986,brinchmann_2004, wyder_2007}.  For SED
fitting, most of the weight on the derived sSFR comes from the UV
to optical ratio.  \cite{salim_2005} show that the sSFR can be
constrained on the basis of the restframe $NUV-r$ color for star
forming galaxies. Alternatively, the restframe $u-r$ color is a
similar, but less precise, sSFR tracer
 \citep[see e.g.,][and references
therein]{wyder_2007}. More generally, fitting to the full SED is
necessary to break degeneracies between age, metallicity and dust.

The measurement of host-galaxy photometry at the SN location is
a direct by-product of all SN~Ia analysis pipelines, because it
must be subtracted from the SN when constructing a lightcurve. For long
duration surveys, most images lack SN light and so can be used
directly, as done here when measuring the local stellar mass.

To measure local sSFR, good spatial resolution is required. The
optimal size for the ``local'' aperture is yet to be determined.
Here and in \cite{rigault_2013} we used a 1~kpc aperture, while in
\cite{rigault_2015} we used a 2~kpc aperture and also found little
change when varying the aperture radius between 1 and 3~kpc.
\cite{kelly_2015} obtain good results using a 5~kpc radius. The
mean seeing values for the SNLS, SDSS, Pan-STARRS and DES deep
imaging are 0.9, 1.1, 1.2, and 1.6~arcsec (FWHM), respectively
\citep[e.g.,][]{guy_2010,jiang_2014,rest_2014,kessler_2015}, so they
have resolution sufficient to extract usable local sSFR measurements
within $\sim{}$3--4~kpc radius apertures out to their highest
redshifts.  Since the angular diameter distance flattens (and
eventually turns over) above $z\sim1$, deep ground-based surveys
going to even higher redshift will have reasonable resolution. Since
ground-based seeing will fluctuate below the median half the time,
the affects of resolution can be examined in detail.

Existing $ugriz$ imaging data for the SNLS, SDSS SN surveys covers
the restframe $u$ through $r$ bands for the redshift range of their
SNe~Ia, and are quite deep. Pan-STARRS and DES lack $u$. Above $z\sim0.4$,
where the SNLS dataset is concentrated, the restframe NUV becomes
accessible via the observer-frame $u$ filter from the CFHTLS Deep
survey. According to \cite{jiang_2014}, the coadded SDSS data in
the ``Stripe 82'' SN search region is about 2 magnitudes deeper
than the main SDSS survey data primarily used in the analysis here.
The SNLS/CFHTLS Deep imaging is about 4~mag deeper than most of the
images used here \citep{hudelot_2012}.  As discussed above, seeing
limitations will necessitate larger local metric apertures, and
this will further improve the signal-to-noise achievable on LsSFR
by high-redshift SN surveys. For instance, in a 3~kpc aperture the
Stripe~82 coadds and the CFHTLS Deep coadds will have sensitivity
out $z\sim0.4$ and $z\sim0.6$, respectively, that is comparable to
that achieved here in a 1~kpc aperture.

Looking ahead to future SN surveys, LSST is expected to have a
median seeing of $\sim 0.7$~arcsec, enabling use of a 3~kpc local
aperture out to $z\sim1$. In addition, thanks to its $u$
filter, LSST will have access to the restframe NUV for SNe~Ia with
redshift $z>0.4$. The final coadds of the LSST Deep Drilling fields
should produce data with similar sensitivity in a 3~kpc as achieved
here in a 1~kpc for SNe~Ia hosts out to $z\sim0.7$. \Instru{Euclid} and
\Instru{WFIRST} images will provide redder filters, reaching the restframe
UV only for the very highest redshifts. But this red coverage will
improve the fitted constraints on sSFR in tandem with ground-based
optical imaging. The \Instru{WFIRST} integral field channel will span
0.4--2~$\mu$m, allowing restframe UV measurements and a wide
wavelength coverage for SED fitting. The space-based data should
go especially deep due to the much darker sky background.
In addition, the full WFIRST program, planned with both
  photometry and integral field spectrophotometry,
  should be able to reproduce the
  LsSFR analysis presented in this paper throughout the planned
  redshift range.

Thus, at this level of detail it seems fair to say that current and
future SN surveys should have the ability to account for the LsSFR
bias.

Even though it appears that LsSFR is measurable from high-redshift
SN survey data, there still will be some practical difficulties.
First, in real SN surveys, signal-to-noise cuts come into play.
Lower stretch SNe~Ia are intrinsically fainter and preferentially
arise in older populations, as shown in
Section~\ref{sec:result_lssfr_stretch}. The greater star formation
activity at higher redshifts leads to more dust and thus more
extinguished SNe~Ia, preferentially in the younger population.
Perhaps most relevant for measuring the LsSFR bias, at a fixed
lightcurve stretch and color, younger SNe are intrinsically fainter,
though in the mean their higher stretch compensates for this.  Thus,
signal-to-noise cuts could suppress either population in ways that
need to be carefully modeled for individual surveys and that could
be very dependent on the actual behavior of the bias. In addition,
as SN surveys become larger, many have come to depend on photometric
classification of transients in lieu of the spectroscopically-classified
SNe used here. Thus, for these surveys the LsSFR bias will need to
be determined in tandem with the effects of photometric classification
error. A recent comparison of host-galaxy correlations with SN
brightnesses found different relations for spectroscopically-classified
and photometrically-classified subsets \citep{wolf_2016}. Future
SN cosmology fitters will need to know of and parameterize many
systematics simultaneously in order to produce unbiased results.
As shown here for LsSFR, the very high-fidelity measurements possible
at low redshift are key to developing such parameterizations
of astrophysical systematics.

\section{Summary and conclusions}
\label{sec:conclusion}

Using a large sample of SNe~Ia from the Nearby Supernova Factory
we have developed and quantified the importance of an improved local
host environment indicator -- the local specific star formation
rate, LsSFR.  Our sample of \nSNe\ SNe~Ia is almost twice that available in
\cite{rigault_2013}.  We derived the local star formation rate from
spatially-resolved H$\alpha$ emission using the methods initially
developed in \cite{rigault_2013}, and the local and global stellar
masses using SDSS and SNIFS $g$- and $i$-band imaging.

LsSFR traces the fraction of young to old stars in the projected
1~kpc radius region around each of our SNe~Ia.  By construction,
this parameter has reduced sensitivity to dust extinction, and we
find only a modest correlation with global gas-phase metallicities for the
subset of our SNe~Ia with metallicity measurements.  We connect
LsSFR with the observed grouping of the SN~Ia delay time distribution
into younger/prompt and older/delayed subpopulations.  We
then use LsSFR to segregate our SN~Ia sample into these younger and
older subpopulations and 
analyze the difference in their standardization properties. Our
results are the following:

\paragraph{\textbf{Lightcurve parameter:}} 
Lightcurve stretch is correlated with LsSFR, with a significance
of $6.5\,\sigma$.  This is in agreement with previous studies
based on other age metrics \citep[e.g.,][]{hamuy_1996,
sullivan_2010,rigault_2013}.  The younger SNe~Ia mainly populate
$x_1>0$ and are more homogeneous in stretch, as shown by their
significantly smaller dispersion in $x_1$.  In contrast, the older
population exhibits a relatively flat distribution over the entire
$-3 < x_1 < 2$ range.  The lightcurve color, on the other hand, has
an insignificant ($\sim 1.5\,\sigma$) correlation with LsSFR.
Thus, we find no evidence for differences in SN~Ia color with
progenitor age.

\paragraph{\textbf{LsSFR-dependent brightness bias:}} 
After performing a conventional linear standardization using \salt{}
stretch and color, we find that SNe~Ia with higher LsSFR are
$\DY = 0.125 \pm 0.023\, \mathrm{mag}$ fainter those with
lower LsSFR. The offset increases to $\DY = 0.163 \pm 0.029\,\mathrm{mag}$
when solving for $\DY$ in the standardization fit, a $\sim6\,\sigma$ result.
Including LsSFR to fit for $\DY$ leads to a significantly
reduced dispersion of $\mathrm{wRMS}=0.129 \pm 0.008\,\mathrm{mag}$.
We have tested that this result is robust against changes in our analysis
and is not the result of overfitting.

\paragraph{\textbf{Standardization by LsSFR subpopulation}} 
We performed independent standardization of the young/prompt and 
old/delayed
SNe~Ia to compare their $\alpha$ and $\beta$ standardization
parameters, finding  that values of $\alpha$ and $\beta$ are
consistent.

When standardizing the younger SNe~Ia alone we find $\mathrm{wRMS}=0.126
\pm 0.010\,\mathrm{mag}$ which is further reduced to $\mathrm{wRMS}=0.120
\pm 0.010\,\mathrm{mag}$ after outlier rejection accounting for 
misattribution of local environment. This confirms previous suggestions that
younger SN~Ia are a more homogeneous population and can provide
distance measurements that are more accurate \citep{rigault_2013,
childress_2014, kelly_2015}.

\paragraph{\textbf{Global stellar mass bias and LsSFR connection}:} 
We find a brightness step of $\Delta_M=0.119\pm0.026\,\mathrm{mag}$
when segregating our sample by host-galaxy global stellar mass.
However, when fitting for LsSFR and global stellar mass biases simultaneously, we
find that $\DM=0.064\pm0.029$ and the LsSFR bias is
$\DY=0.129\pm0.032$. 
The reduction in $\DM$ when including LsSFR is
significant at greater than $5\,\sigma$.  We therefore conclude
that the stellar mass bias is, at least partially, caused by the
LsSFR/age bias, as originally suggested by \cite{rigault_2013} and
modeled by \cite{childress_2014}.

The strength of $\DY$ relative to $\DM$ indicates
that including only $\DM$ in cosmological analyses does
remove redshift-dependent or sample-selection bias from the fitted
dark energy parameters. But also, although LsSFR account for most of
the step, since we find $\DM$ to be
detected at $2.2\,\sigma$, it is possible that it encodes another
astrophysical bias not captured by LsSFR.  The complex origin of
the stellar mass bias further emphasizes the difficulty of using
such a poorly controlled indicator for SN~Ia cosmology.
 
\paragraph{\textbf{Impact on cosmology}:} 
The ratio between the fraction of SNe~Ia from younger or older
progenitors follows -- by definition -- the steep $(1+z)^{2.8}$
\citep{tasca_2015} evolution of sSFR in the universe. We have
simulated the plausible impact of the LsSFR bias on the derivation
of the dark energy equation of state parameters $w_0$ and $w_a$.
We find that $w_0$ is shifted toward lower values ($-0.03$ including
$z>1$ SNe~Ia and $-0.05$ without). The greatest impact is on $w_a$,
which has a strong negative bias ($-0.7 \pm 0.1$).  Thus the offset
between younger and older SNe~Ia has the ability to bias the
determination of the dark energy equation of state parameters very
badly if not taken into account.

\paragraph{\textbf{Measuring LsSFR for other SN~Ia cosmology surveys}:} 
LsSFR can be measured in several ways, and for long-running high-redshift
surveys that build up deep imaging it can be obtained from SED
fitting, especially because restframe UV coverage is usually
available.  Current SN~Ia samples like SDSS and SNLS have already
obtained optical data that can provide the local galaxy information
needed to assess which SNe~Ia are likely to be younger or older. In
the future, the increased depth and better angular resolution
expected from the LSST, \Instru{Euclid}, and \Instru{WFIRST} SN~Ia surveys can obtain
this information out to even higher redshifts.

In conclusion, the locally-measured specific star formation rate appears
able to segregate SNe~Ia by age, and doing so is one of the most
essential ingredients in obtaining unbiased cosmological results
from SNe~Ia.

\begin{acknowledgements}

We thank the technical staff of the University of Hawaii 2.2 m
telescope, and Dan Birchall for observing assistance. We recognize
the significant cultural role of Mauna Kea within the indigenous
Hawaiian community, and we appreciate the opportunity to conduct
observations from this revered site. This project has received funding
from the European Research Council (ERC) under the European Union’s
Horizon 2020 research and innovation programme (grant agreement No
759194 — USNAC). This work was supported in
part by the Director, Office of Science, Office of High Energy
Physics of the U.S. Department of Energy under Contract No. DE-AC02-
05CH11231.  Support in France was provided by CNRS/IN2P3, CNRS/INSU,
and PNC; LPNHE acknowledges support from LABEX ILP, supported by
French state funds managed by the ANR within the Investissements
d'Avenir programme under reference ANR-11-IDEX-0004-02.  NC is
grateful to the LABEX Lyon Institute of Origins (ANR-10-LABX-0066)
of the University de Lyon for its financial support within the
program "Investissements d'Avenir" (ANR-11-IDEX-0007) of the French
government operated by the National Research Agency (ANR).  Support
in Germany was provided by DFG through TRR33 "The Dark Universe"
and by DLR through grants FKZ 50OR1503 and FKZ 50OR1602.  In China
the support was provided from Tsinghua University 985 grant and
NSFC grant No  11173017.  Some results were obtained using resources
and support from the National Energy Research Scientific Computing
Center, supported by the Director, Office of Science, Office of
Advanced Scientific Computing Research of the U.S. Department of
Energy under Contract No. DE-AC02- 05CH11231. We thank the Gordon
\& Betty Moore Foundation for their continuing support. Additional
support was provided by NASA under the Astrophysics Data Analysis
Program grant 15-ADAP15-0256 (PI: Aldering). We also thank the High
Performance Research and Education Network (HPWREN), supported by
National Science Foundation Grant Nos. 0087344 \& 0426879.  

Some SN discovery observations were obtained with the Samuel Oschin
Telescope at the Palomar Observatory as part of the Palomar Transient
Factory project, a scientific collaboration between the California
Institute of Technology, Columbia University, Las Cumbres Observatory,
the Lawrence Berkeley National Laboratory, the National Energy
Research Scientific Computing Center, the University of Oxford, and
the Weizmann Institute of Science.

This research has made use of the NASA/IPAC Extragalactic Database
(NED), which is operated by the Jet Propulsion Laboratory, California
Institute of Technology, under contract with the National Aeronautics
and Space Administration.

Funding for the SDSS~I--III has been provided by the Alfred P. Sloan
Foundation, the Participating Institutions, the National Science
Foundation, the U.S. Department of Energy, the National Aeronautics
and Space Administration, the Japanese Monbukagakusho, the Max
Planck Society, and the Higher Education Funding Council for England.
The SDSS Web Site is http://www.sdss.org/.  The SDSS is managed by
the Astrophysical Research Consortium for the Participating
Institutions. The Participating Institutions for SDSS and SDSS-II
are the American Museum of Natural History, Astrophysical Institute
Potsdam, University of Basel, University of Cambridge, Case Western
Reserve University, University of Chicago, Drexel University,
Fermilab, the Institute for Advanced Study, the Japan Participation
Group, Johns Hopkins University, the Joint Institute for Nuclear
Astrophysics, the Kavli Institute for Particle Astrophysics and
Cosmology, the Korean Scientist Group, the Chinese Academy of
Sciences (LAMOST), Los Alamos National Laboratory, the Max-Planck-Institute
for Astronomy (MPIA), the Max-Planck-Institute for Astrophysics
(MPA), New Mexico State University, Ohio State University, University
of Pittsburgh, University of Portsmouth, Princeton University, the
United States Naval Observatory, and the University of Washington.
For SDSS-III they are the University of Arizona, the Brazilian
Participation Group, Brookhaven National Laboratory, Carnegie Mellon
University, University of Florida, the French Participation Group,
the German Participation Group, Harvard University, the Instituto
de Astrofisica de Canarias, the Michigan State/Notre Dame/JINA
Participation Group, Johns Hopkins University, Lawrence Berkeley
National Laboratory, MPIA, Max Planck Institute for Extraterrestrial
Physics, New Mexico State University, New York University, Ohio
State University, Pennsylvania State University, University of
Portsmouth, Princeton University, the Spanish Participation Group,
University of Tokyo, University of Utah, Vanderbilt University,
University of Virginia, University of Washington, and Yale University.

\end{acknowledgements}

\begin{appendix}

\section{Step measurements}
\label{app:fitstep}
To derive the LsSFR or global stellar mass step values, we use the sum of two
normal distributions ($a$ and $b$) with mean brightnesses and standard deviations
$\mu_a$, $\,\sigma_a$ and $\mu_b$, $\sigma_b$, respectively, to represent
the underlying parent populations. Each data point
has a probability $p_i$ of being associated with mode $a$ and a
probability $1-p_i$ of being associated with mode $b$. 
Each datum has a measurement uncertainty $\sigma_i$ and an observed value $x_i$. The
likelihood, $\mathcal{L}_i$, of observing $x_i$, given the bi-normal model
and the measurement uncertainties $\sigma_i$, is:
\begin{equation}
\label{eq:step}
\mathcal{L}_i = p_i \times \frac{1}{\sqrt{2\pi (\sigma_a^2 +
  \sigma_i^2)}} \exp\left(\frac{ - (\mu_a-x_i)^2}{2 (\sigma_a^2 +
  \sigma_i^2)}\right) + (1-p_i) \times \frac{1}{\sqrt{2\pi (\sigma_b^2 +
  \sigma_i^2)}} \exp\left(\frac{ - (\mu_b-x_i)^2}{2 (\sigma_b^2 +
  \sigma_i^2)}\right)
\end{equation}

We then minimize $- \sum_i \log(\mathcal{L}_i)$ to extract the
mean brightness and dispersion of each mode. The quoted brightness step is the
difference between the means, and the step uncertainty is the quadrature sum
of the fitted uncertainties on the means. We used MCMC to confirm that the two
means are uncorrelated.

The \salt{} algorithm returns an irreducible uncertainty of approximately
0.011 in $x_1$ and 0.018~mag in $c$ in order to account for unexplained
scatter when training its model. We remove these uncertainty floors
from the $\sigma_i$ so they are absorbed into $\sigma_a$ and $\sigma_b$,
where they belong.  For typical standardization coefficients, this
irreducible \salt{} ``intrinsic dispersion'' is around 0.055~mag.

We use this method when fitting for Hubble residual steps, i,e, on
data that have already been standardized using a linear stretch and
color correction.

\end{appendix}

\end{document}